\documentclass[a4paper,12pt]{article}
\pdfoutput=1 

\usepackage{jheppub} 

\usepackage[T1]{fontenc} 

\usepackage{amsmath}
\usepackage{dsfont}
\usepackage{xcolor}

\def\be{\begin{equation}}
\def\ee{\end{equation}}
\def\n{\nonumber \\}

\def\bea{\begin{eqnarray}}
\def\eea{\end{eqnarray}}
\def\t{\tilde}

\title{\boldmath Tensionless strings on ${\rm AdS}_3$ orbifolds}

\author[a]{Matthias R.\ Gaberdiel,}
\author[b]{Bin Guo,}
\author[c]{and Samir D.\ Mathur}

\affiliation[a]{Institut f\"ur Theoretische Physik, ETH Zurich, \\ CH-8093 Z\"urich, Switzerland}
\affiliation[b]{Institut de Physique Th\'eorique, Universit\'e Paris-Saclay, \\
CNRS, CEA, Orme des Merisiers, Gif-sur-Yvette, 91191 CEDEX, France}
\affiliation[c]{Department of Physics, The Ohio State University, \\ Columbus, OH 43210, USA}

\emailAdd{gaberdiel@itp.phys.ethz.ch}
\emailAdd{bin.guo@ipht.fr}
\emailAdd{mathur.16@osu.edu}

\abstract{The bound state of one NS5 brane (wrapped on a $\mathbb{T}^4$) and $N$ NS1-branes has two dual descriptions: its low-energy dynamics is described by the symmetric orbifold of $\mathbb{T}^4$, while the near horizon geometry is captured by string theory on ${\rm AdS}_3 \times {\rm S}^3\times \mathbb{T}^4$ with one unit of NS flux. The latter theory is exactly solvable in the hybrid formalism, and this allows one to prove the equivalence of the two descriptions. In this paper we extend this duality to $\mathbb{Z}_k$ orbifolds of this ${\rm AdS}_3 \times {\rm S}^3$ background. In particular, we show that the corresponding worldsheet spectrum reproduces exactly the perturbative excitations on top of  a certain non-perturbative state in the dual symmetric orbifold theory. Since the ${\rm AdS}/{\rm CFT}$ duality map is exact for these models, we obtain an interesting picture of how the duality relates boundary and bulk descriptions.}

\begin{document} 
\maketitle
\flushbottom

\section{Introduction}

 String theory gives rise to a consistent theory of quantum gravity, but it is not an easy theory to study.  The gauge-gravity correspondence offers a way to recast string theory as a quantum field theory without gravity. However, this correspondence relates gravity at weak coupling to a field theory at strong coupling, making the study of the field theory difficult as well.  

A little while ago \cite{Gaberdiel:2018rqv,Eberhardt:2018ouy,Eberhardt:2019ywk}, an exact string theory dual was found for a field theory at {\it weak} coupling. The string theory lives on ${\rm AdS}_3\times {\rm S}^3\times {\mathcal M}$, where ${\mathcal M}$ is $\mathbb{T}^4$ or K3. This string theory can be described on the string worldsheet by a supersymmetric WZW model at level ${\sf k}=1$ for the supergroup ${\rm PSU}(1,1|2)$, using the hybrid formalism of \cite{Berkovits:1999im}. The dual field theory is a  2-dimensional CFT,  described by a sigma model.  The target space of this sigma model is   an orbifold ${\mathcal M}^N/S_N$, the symmetric product of $N$ copies of the 4-manifold ${\mathcal M}$, in the limit $N\rightarrow\infty$. 

The state of string theory given by global ${\rm AdS}_3 \times {\rm S}^3 \times {\mathcal M}$ is dual to the vacuum state of the dual CFT in the `untwisted' sector. Excitations above this vacuum state of the dual CFT map to `long strings' in the string theory. Thus one can study the ${\rm AdS}/{\rm CFT}$ map exactly for all states that can be obtained perturbatively from this vacuum state of the dual CFT.
\medskip

In the present paper, we will extend this ${\rm AdS}/{\rm CFT}$ map to a specific class of states that are perturbative excitations above a new non-perturbative ground state. More specifically, we will consider the  states in the symmetric orbifold where all $N$ copies are arranged into $N/k$ $k$-cycles. This multi-particle state has infinite energy relative to the original ground state of the symmetric orbifold theory in the $N\rightarrow\infty$ limit; it therefore describes a `non-perturbative' new vacuum.  

The corresponding spacetime string background is an orbifold of ${\rm AdS}_3\times {\rm S}^3$ by $\mathbb{Z}_k$, times ${\mathcal M}$, in agreement with the old proposal of \cite{Martinec:2001cf}.\footnote{We will use the `sans-serif' font for the level ${\sf k}$ to distinguish it from the order of the orbifold group $k$. For most of the paper will consider the case ${\sf k}=1$. Our analysis also fits with the general result of \cite{Eberhardt:2020bgq,Eberhardt:2021jvj}.} These orbifolds can be described, in the worldsheet theory, as orbifolds of the level ${\sf k}=1$ $\mathfrak{psu}(1,1|2)_1$ WZW model. We shall check this proposal by confirming that the perturbative single-particle excitation spectrum above these new ground states is precisely reproduced from the worldsheet. Among other things our analysis therefore shows that the orbifold background can be regarded as a condensation of the $k$-cycle twisted sector states. 
\medskip

In a nice set of papers \cite{Martinec:2001cf,Martinec:2002xq,Martinec:2017ztd,Martinec:2019wzw,Martinec:2018nco,Martinec:2020gkv,Bufalini:2021ndn,Martinec:2022okx,Martinec:2023zha}, the worldsheet string theory for ${\rm AdS}_3\times {\rm S}^3\times \mathbb{T}^4$ and its orbifolds was studied for the case where the geometry was created by a large number of NS1 branes bound to $\sf k$ NS5 branes. In their approach, which builds on the pioneering work of  \cite{Maldacena:2000hw}, the number of NS5 branes satisfies ${\sf k}\ge 2$, while the dual symmetric orbifold theory is now believed to lie at ${\sf k}=1$. The difference in ${\sf k}$ does not matter for the BPS states of the theory, but the full (non-BPS) spectrum depends on ${\sf k}$, and thus it will be important to consider the worldsheet theory at ${\sf k}=1$ if one wants to match the full (non-BPS) excitation spectrum, as we will do.\footnote{Note that by using a set of discrete string dualities (and continuous deformations of moduli) one can connect the situation of $\sf k$ NS5 branes and $n_1$  NS1 branes to the case of ${\sf k'}=1$ NS5 brane and $n_1'={\sf k}n_1$ NS1 branes. But under these dualities, the perturbative string modes in one set-up map to higher dimensional branes wrapping various cycles of the $\mathbb{T}^4$ in the other. Thus there is no simple map between the worldsheet theories for ${\sf k}=1$ and for ${\sf k}\ge 2$.} In turn, our exact correspondence will give us insights into the string theoretic description of the geometries that are associated to the individual symmetric orbifold states, thus going beyond what was possible in \cite{Martinec:2001cf}. On the other hand, our results are in broad agreement with what was found there: in particular, the spacetime geometry that is dual to the specific non-perturbative symmetric orbifold states turns out to match precisely with what was predicted in \cite{Martinec:2001cf}.\footnote{For other interesting studies of the AdS/CFT duality for the NS1-NS5 brane system, see \cite{Giveon:2005mi,Dei:2021xgh,Dei:2021yom,Balthazar:2021xeh, Martinec:2021vpk,Eberhardt:2021vsx,Dei:2022pkr}.}

\subsection{Summary of results}

Let us now describe in a little more detail the main results of our paper. 
As we mentioned before, the dual CFT is described by $N$ copies of a $c=6$ CFT living on a spatial circle of length $2\pi$ (modulo the identification of the copies by the symmetric group $S_N$ and considering the large $N$ limit). The state where, say, $k$ of these copies are joined together to form a `multi-wound' string living on a circle of length $2\pi k$, describes a perturbative excitation of the symmetric orbifold ground state, and it corresponds to a perturbative world-sheet state in the above hybrid string description. 
\bigskip

(1) In this paper we shall be interested in the non-perturbative excitations where {\it all} $N$ copies of the $c=6$ CFT are joined together in $N/k$ sets of $k$. These states do not describe a perturbative excitation since the difference in energy to the ground state is proportional to $N$, the total number of copies, and therefore goes to infinity in the large $N$ limit. We can furthermore look at the perturbative excitations relative to this background of $k$-wound strings. The main question we want to address is: what is the string theory description of this non-perturbative symmetric orbifold state and its perturbative excitations? We shall find very convincing evidence that the corresponding string worlsheet theory is obtained by taking an orbifold of the the $\mathfrak{psu}(1,1|2)_1$ WZW theory by an orbifold group of order $k$. 
\bigskip

(2) We shall also analyse this problem in terms of the dual geometry, and the specific dictionary we shall find is in fact suggested by a careful analysis of the coordinates in the AdS geometry and their relation to the dual CFT. More specifically, there are different choices of what we may want to regard as the time coordinate in spacetime, and they will lead to different definitions of the energy $E_{{\rm ws}}$ of a string excitation as computed from the string worldsheet theory. We can also choose different time coordinates in the dual CFT, and this will lead to different definitions for the energy $E_{{\rm CFT}}$. 

As we shall explain, there is a particular choice of the time coordinate that is physically important. Recall that the dual CFT arises from the low energy dynamics of a bound state of NS1 and NS5 branes. Starting with type IIB string theory compactified on ${\rm S}^1\times \mathbb{T}^4$, the NS1 branes wrap the $S^1$, while  the NS5 branes wrap ${\rm S}^1\times \mathbb{T}^4$. The fermions in the 10-dimensional string theory are periodic around the ${\rm S}^1$, and thus the induced fermionic degrees of freedom on the NS1-NS5 bound state are also periodic around the ${\rm S}^1$.  Thus the dual CFT is naturally described in the Ramond sector, where the fermions are periodic around the spatial circle of the CFT. Similarly, the time coordinate that is natural to the dual CFT is the time induced on the branes from the time coordinate of the 10-dimensional string theory; this is the time measured by observers far away who may wish to scatter quanta off the NS1-NS5 bound state.

On the other hand, in the bulk gravity theory, a natural choice of coordinates is the one where the directions associated with the ${\rm S}^3$ are orthogonal to those corresponding to ${\rm AdS}_3$. In fact, in our worldsheet description we are implicitly using such a choice of coordinates since the bosonic subalgebra of $\mathfrak{psu}(1,1|2)$ is the direct sum of $\mathfrak{su}(2)\oplus \mathfrak{sl}(2,\mathds{R})$, where $\mathfrak{su}(2)$ and  $\mathfrak{sl}(2,\mathds{R})$ correspond to ${\rm S}^3$ and  ${\rm AdS}_3$, respectively. It turns out that these two `natural' choices of coordinates --- the one where time is induced from the description at infinity, vs the one where ${\rm S}^3$ and ${\rm AdS}_3$ are orthogonal --- are not the same, but there is a simple map between them, which we will describe. We shall find that, using the corresponding map between the energies in the dual CFT and the string theory, leads to a perfect match between the worldsheet theory and the dual CFT description. 
\bigskip

(3)  The bound state of NS1 and NS5 branes describes a 2-charge object, which gives rise to the microstates of a `small black hole'. In order to account for the microstates of the 3-charge NS1-NS5-P system which were famously used to study the 3-charge black hole  entropy \cite{Strominger:1996sh}, one needs to add momentum charge to the system. In fact, if we add excitations that travel in both directions along the $S^1$,   we get in general non-extremal states in the NS1-NS5-P-${\bar {\rm P}}$ system, which are expected to be dual to non-extremal black holes.

As we shall explain, the considerations above can also be generalised to that set-up. From the worldsheet perspective, this amounts to considering an {\it asymmetric} $\mathbb{Z}_k$ orbifold of the $\mathfrak{psu}(1,1|2)_1$ WZW model, i.e.\ the orbifold acts differently on left- and right-movers on the world-sheet. The dictionary explained in (2) then also allows us to identify the corresponding states in the dual CFT. In fact, given that the worldsheet theory is now an asymmetric orbifold, there is a non-trivial constraint, the so-called level matching condition \cite{Narain:1986qm}, that needs to be respected in order for the worldsheet theory to be well-defined. We will find that this constraint has a direct analogue in the symmetric orbifold, namely that it corresponds to the requirement that the relevant dual CFT ground states are orbifold invariant.

\subsection{Plan of the paper}

The paper is organised as follows. The duality between string theory on ${\rm AdS}_3\times {\rm S}^3 \times \mathbb{T}^4$ with one unit (${\sf k}=1$) of NS-NS flux and the dual symmetric orbifold theory is reviewed in detail in Section~\ref{sec:review}. In Section~\ref{sec orbifold} we study the symmetric $\mathbb{Z}_k$ orbifold of the ${\sf k}=1$ worldsheet theory, and show that its spacetime spectrum reproduces the expected states in the symmetric orbifold theory. The generalisation to the asymmetric orbifold construction is described in Section~\ref{sec:genorb}. Section~\ref{sec:discussion} contains our conclusions, and there are a number of appendices where some of the more technical material can be found.

\section{Review of the  spectrum on ${\rm AdS}_3\times {\rm S}^3 \times \mathbb{T}^4$}\label{sec:review}

In this section we give a brief review of the spectrum of string theory on ${\rm AdS}_3\times {\rm S}^3 \times \mathbb{T}^4$ with one unit of NS-NS flux. We also review the dual symmetric orbifold CFT and its spectrum. 

\subsection{The geometry of ${\rm AdS}_3\times {\rm S}^3 \times \mathbb{T}^4$}\label{secgeometries}

Our analysis will be for the case where the number of NS5 branes is ${\sf k}=1$. For this small number of NS5 branes, the string theory spacetime in the near-brane region is not well-described by classical geometry. Nevertheless, we can get some intuition for the kind of orbifolds that we expect by examining the orbifold geometries that arise for large numbers of NS5 and NS1 branes. Thus
 let us begin by reviewing the geometry ${\rm AdS}_3\times {\rm S}^3 \times \mathbb{T}^4$ which arises as the near horizon geometry of the NS1-NS5 bound state. More specifically, we start with type IIB string theory  compactified on ${\rm S}^1 \times \mathbb{T}^4$, and wrap $n_5$ NS5 branes on ${\rm S}^1 \times \mathbb{T}^4$, and $n_1$ NS1 branes on $S^1$. The bound state of these branes has many degenerate extremal solutions, which give rise to the entropy $S=2\sqrt{2}\sqrt{n_1n_5}$. Each state corresponds to a different gravity solution, and such solutions were found in \cite{Lunin:2002bj}. The solution with largest angular momentum is particularly simple. In the near-horizon limit, the geometry in the string frame is
\bea
ds^2_{\rm string}&=&{f\over \sqrt{Q_1Q_5}}\left[-d\tilde{t}^{\, 2}+d\t y^2\right]+\sqrt{Q_1Q_5}\Big[{d\t r^2\over \t r^2+a^2}+d\t \theta^2+\cos^2\!\t \theta d\t \psi^2+\sin^2\!\t \theta d\t \phi^2\Big]\nonumber\\
&& -\, 2a \sin^2\!\t\theta d\t\phi d\t t-2a\cos^2\!\t\theta d\t\psi d\t y \ ,
\label{mmfive0}
\eea
with
\be
f=\t r^2+a^2\cos^2\!\t \theta \ , ~~~a={\sqrt{Q_1Q_5}\over R_y} \ .
\ee
Here 
\be
Q_1={g^2 \alpha'^3\over V} n_1 \ , ~~~Q_5=\alpha' n_5 \ ,
\ee
where $R_y$ is the radius of the $\rm S^1$, and $(2\pi)^4 V$ is the volume of the $\mathbb{T}^4$. 
We write $Q=\sqrt{Q_1Q_5}$, and rescale the coordinates via 
\be\label{rescale1}
t_R={\t t\over R_y} \ , ~~~y_R={\t y\over R_y} \ , ~~~r_R={\t r\over a} \ , ~~~\theta_R=\t \theta \ , ~~~\psi_R=\t\psi \ , ~~~\phi_R=\t\phi \ ,
\ee
to bring the metric into the form 
\bea
ds^2_{\rm string}&=&Q\bigg [(r_R^2+\cos^2\!\theta_R)(-dt_R^2+dy_R^2) \nonumber \\
& & \qquad + \Big({dr_R^2\over r_R^2+{1}}+d\theta_R^2+\cos^2\!\theta_R d\psi_R^2+\sin^2\!\theta_R d\phi_R^2\Big)\\
&&\qquad \ - 2 \sin^2\!\theta_R d\phi_R dt_R- 2\cos^2\!\theta_R d\psi_R dy_R \bigg ]
\label{mmtwo} \ . \nonumber
\eea
Here the subscript $R$ refers to the fact that the boundary CFT dual to the geometry in these coordinates will be a CFT in the {\bf Ramond sector}. Recall that if we continue the near-region geometry (\ref{mmfive0}) to larger $r$ we go over to asymptotically flat spacetime. The compact ${\rm S}^1$ that is parametrized by $y_R$ stabilizes to a constant radius at spatial infinity, and the fermions must be periodic around this ${\rm S}^1$ in order to avoid a cosmological constant in this asymptotically flat region. 

The metric (\ref{mmtwo}) has off-diagonal terms, which can be removed by a coordinate transformation. To this end we define
\be
t=t_R \ , ~~~y=y_R \ , ~~~r=r_R \ , ~~~\theta=\theta_R \ , ~~~\psi = \psi_R -   y_R \ , ~~~ \phi = \phi _R -   t_R \ ,
\label{mmthree}
\ee
so that the metric becomes
\be
ds^2_{\rm string}=Q\big[-(r^2 + 1) dt^2 + {d r^2\over  r^2 +1} +  r^2 dy^2 + d \theta^2 +\cos^2\!\theta d \psi^2 +\sin^2\!\theta d \phi^2\big] \ .
\label{mmone}
\ee
This is the metric of global ${\rm AdS}_3\times {\rm S}^3$. 

In order to understand the identifications recall that the Ramond sector coordinates are simply related to those that describe flat spacetime at spatial infinity, see eq.~(\ref{rescale1}), and hence satisfy the identifications
\be
(y_R, \psi_R, \phi_R)\sim (y_R+2\pi, \psi_R, \phi_R)\sim (y_R, \psi_R+2\pi, \phi_R)\sim (y_R, \psi_R, \phi_R+2\pi) \ .
\ee
Given (\ref{mmthree}), the identification of our final coordinates is therefore 
\be
(y,\psi, \phi) \sim (y+2\pi, \psi -2\pi , \phi)\sim
 (y, \psi +2\pi , \phi )
 \sim (y, \psi  , \phi +2\pi) \ ,
\ee
and hence each of the coordinates $y,  \psi, \phi$ is simply periodic with period $2\pi$. As regards the boundary conditions for the fermions, in the metric (\ref{mmone}) the $y$ circle shrinks to zero size smoothly at $r=0$. Thus the fermions need to be antiperiodic around the $y$ circle in (\ref{mmone}), in order to avoid a singularity at $r=0$. This is consistent with the fact that the fermions carry charge $1/2$ under the rotations of the ${\rm S}^3$, and thus change their periodicity under the coordinate transformation (\ref{mmthree}). The coordinates (\ref{mmone}) are therefore the coordinates that are dual to a CFT in the {\bf Neveu-Schwarz (NS) sector}.

\subsection{Worldsheet analysis}\label{sec worldsheet}

Let us next turn to the worldsheet description of this background. For that it is helpful to first recall covariant bosonic string theory in flat space and then on ${\rm AdS}_3$. 

\subsubsection{The bosonic case}

The Fock space of worldsheet degrees of freedom (of bosonic string theory) in flat space consists of the vectors\footnote{In the following we shall mainly be concentrating on the left-moving degrees of freedom; the analysis for the right-movers is similar.}
\be
|\psi\rangle = \alpha^{\mu_1}_{-n_1} \cdots \alpha^{\mu_l}_{-n_l}\, |p\rangle \ , 
\ee
where $|p\rangle$ describes a momentum ground state with momentum $\alpha^\mu_0 \, |p\rangle = p^\mu \, |p\rangle$, while the $\alpha^{\mu_i}_{-n_i}$ are the creation oscillators. (The positive modes $\alpha^{\mu}_n$ with $n>0$ annihilate the ground state $|p\rangle$.) Note that the different ground states $|p\rangle$ transform under the action of the Lorentz group, the symmetry group of Minkowski space. The physical states of the string theory are then selected by the condition that 
\bea\label{stringcond}
L_n|\psi\rangle&=&0 \ , ~~~ n>0 \ ,\nonumber\\
(L_0-a)|\psi\rangle&=&0 \ ,
\eea
where $a$ is a normal ordering constant, and the second condition is usually referred to as the `mass-shell condition'. 

The situation for bosonic string theory on ${\rm AdS}_3$ with ${\sf k}$ units of NS flux (and no R flux) is similar \cite{Maldacena:2000hw}. Now the Fock space of states is spanned by the vectors 
\be\label{psiAdS}
|\psi\rangle_{\rm AdS} = J^{a_1}_{-n_1} \cdots J^{a_l}_{-n_l}\, |j,m\rangle_{\rm AdS} \ , 
\ee
where the $J^a_n$ are the modes of the $\mathfrak{sl}(2,\mathds{R})_{\sf k}$ current algebra,
\be\label{sl2R}
\begin{aligned}
{}[ J^3_m,J^3_n] & = - \tfrac{{\sf k}}{2} m \delta_{m,-n} \\
{}[J^3_m,J^\pm_n] & = \pm J^\pm_{m+n} \\
{}[J^+_m,J^-_n] & = - 2 J^3_{m+n} + {\sf k} \, m \delta_{m,-n} \ . 
\end{aligned}
\ee
Furthermore, the ground states $|j,m\rangle$ (that are annihilated by the $J^a_n$ modes with $n>0$) sit in a representation of $\mathfrak{sl}(2,\mathds{R})$, the symmetry algebra of ${\rm AdS}_3$. Here $j$ is the  spin of the $\mathfrak{sl}(2,\mathds{R})$ representation, i.e.\ it labels its Casimir $C=-j(j-1)$, while $m$ denotes the $J^3_0$ eigenvalue and thus parametrises the different states in this representation. (In comparison to the flat space case, we can thus identify $-j(j-1) = p^2$, while $m$ corresponds to the different momentum vectors $p^\mu$ with this fixed value of $p^2$.) For the case of ${\rm AdS}_3$ the representations that actually appear in the worldsheet spectrum \cite{Maldacena:2000hw} are either the so-called discrete representations for which $j$ is real and $m= j, j+1,j+2,\ldots$ with $J^-_0 |j,j\rangle=0$; or the continuous representations for which $j=\tfrac{1}{2} + is $ with $s$ real, and $m$ takes values in $m\in \alpha + \mathbb{Z}$. In fact, all such representations appear, i.e.\ the theory is the `diagonal' modular invariant (combining all these representations but with the condition that the representation for the left- and right-movers is always the same), except that  the spin of the discrete representations is bounded by $\frac{{\sf k}}{2}$, see the discussion at the end of this section.

The physical states in string theory are then again characterised by the condition (\ref{stringcond}). If the background involves, in addition to ${\rm AdS}_3$, some other `internal' space, then the Virasoro generators are the sum of those associated to $\mathfrak{sl}(2,\mathds{R})_{\sf k}$, and those associated to the `internal' space, $L_n=L_n^{\rm AdS}+L^{\rm int}_n$. Given the usual Sugawara construction for $L_n^{\rm AdS}$, the $L_0^{\rm AdS}$ eigenvalue on the state $|\psi\rangle_{\rm AdS}$ in (\ref{psiAdS}) is 
\be\label{L0AdS}
L_0^{\rm AdS} \, |\psi\rangle_{\rm AdS} = \Bigl( \frac{-j(j-1)}{{\sf k}-2} + \underbrace{\sum_{i=1}^{l} n_i}_{N^{\rm AdS}} \Bigr)\, |\psi\rangle_{\rm AdS} \ . 
\ee
Geometrically, the states for which the ground states sit in a discrete representation describe the so-called `short strings' that are localised near the center of AdS, while the states that come from the continuous representations correspond to the so-called `long strings' that come close to the boundary of AdS. In particular, the latter have spin $j=\frac{1}{2} + is$, where the continuous parameter $s$ describes the radial momentum of the long strings.\footnote{${\rm AdS}_3$ with pure NS flux possesses these long string solutions since the tension that would make the strings shrink towards the center of AdS is compensated by the NS flux that stabilises them. As a consequence the radial direction is a flat direction, and the corresponding radial momentum parameter $s$ is therefore continuous.} In either case, in terms of the dual 2d CFT, the 2d conformal dimension is identified with the $J^3_0$ eigenvalue of the physical state $|\psi\rangle_{\rm AdS}$. (In the fermionic case this is slightly more subtle, and is explained in some detail in Section~\ref{sec:iden}.)

The no-ghost theorem for bosonic strings on ${\rm AdS}_3$ \cite{Hwang:1990aq,Evans:1998qu} implies that only the discrete representations with $0<j< \frac{{\sf k}}{2}$ are allowed, i.e.\ these are the only representations for which the no-ghost theorem holds (and that therefore lead to a unitary spacetime spectrum). Given the form of (\ref{L0AdS}) this gives a lower bound on the value of the Casimir --- the flat space analogue would be that $p^2$ is bounded from below --- and hence the mass-shell condition implies that there is an upper bound on the excitation number, i.e.\ on $N^{\rm AdS}$ and the corresponding excitation number from the `internal' CFT. (Since the time direction is already accounted for in terms of the ${\rm AdS}_3$ part, the internal CFT should have a positive spectrum.) As a consequence, it is clear  that the above representations do not suffice to describe string theory on ${\rm AdS}_3$ \cite{Maldacena:2000hw}. 

\subsubsection{Spectral flow}\label{sec:bosspectral}

This puzzle was resolved in a beautiful paper \cite{Maldacena:2000hw} in which the additional representations were identified:\footnote{The same representations had previously been postulated in \cite{Henningson:1991jc} based on the requirement that the worldsheet theory should be modular invariant.} they are obtained from the above highest weight representations by spectral flow (by $w$ units), and they describe string solutions that wind $w$ times around the boundary of ${\rm AdS}_3$. (Note that this winding number $w$ is not topologically protected, i.e.\ the circle along the boundary of ${\rm AdS}_3$ is contractable.  However, these winding solutions are nevertheless stable.) More specifically, given any of the allowed highest weight representations on which the modes of the $\mathfrak{sl}(2,\mathds{R})_{\sf k}$ act as $\tilde{J}^a_n$, we define the $w$-spectrally flowed representation by the `spectrally flowed action' defined by 
\be\label{ggone}
\begin{aligned}
J^3_n  &=   \tilde J^3_n + \tfrac{{\sf k} w}{2} \delta_{n,0} \ , \\
J^\pm_n  &=  \tilde J^\pm_{n\mp w}  \ ,\\
L_n &=   \tilde L_n - w \tilde J^3_n  - \tfrac{\sf k}{4} w^2 \delta_{n,0} \ .
\end{aligned}
\ee
It is easy to check that this defines another representation of $\mathfrak{sl}(2,\mathds{R})_{\sf k}$ -- this just amounts to the statement that the identification in (\ref{ggone}) defines an automorphism of the algebra, i.e.\ that it respects the commutation relations. In terms of the ${\rm AdS}_3$ geometry, spectral flow modifies the 
boundary dependence of a solution by, see \cite[eq.~(26)]{Maldacena:2000hw} with $w=w_L=w_R$
\begin{align}
& t(\tau,\sigma) \mapsto t(\tau, \sigma)  \ , \n
& y(\tau,\sigma) \mapsto y(\tau,\sigma) + w \, \sigma \ , \label{spectral}
\end{align}
where $t$ and $y$ are the coordinates in (\ref{mmone}). In particular, it therefore introduces winding by $w$ units in the $y$ direction. From the algebraic perspective of solving eq.~(\ref{stringcond}) the main effect is that it modifies the mass-shell condition to
\bea
(L_0-a)|\psi\rangle&=& (\t L_0 - w \t J^3_0 -  \tfrac{\sf k}{4} w^2 -a)|\psi\rangle\nonumber \\[2pt]
&=&(\t L^{\rm AdS}_0 + L^{\rm int}_0 - w \t J^3_0 -  \tfrac{\sf k}{4} w^2 -a)|\psi\rangle\nonumber \\
&=& \bigl(-\tfrac{j(j-1)}{{\sf k}-2}+N^{\rm AdS} + h_{0}+ N^{\rm int} -w \t m -  \tfrac{\sf k}{4} w^2 -a \bigr)|\psi\rangle=0  \ ,
\eea
where we have assumed that $L^{\rm int}_0 |\psi\rangle = ( h_{0}+ N^{\rm int}) |\psi\rangle$, where $h_0\geq 0 $ is the internal ground state energy, while $N^{\rm int}\in\mathbb{N}_0$ is the internal excitation number. 
At least for the continuous representations we can solve this easily for $\tilde{m}$ --- recall that $\alpha$ that specifies the $J^3_0$ eigenvalues modulo integers, see the paragraph below eq.~(\ref{sl2R}), is an independent parameter from $j=\frac{1}{2} + is$, 
\be
\t m = {1\over w}\big( - \tfrac{j(j-1)}{{\sf k}-2} +N^{\rm AdS}+ h_0 + N^{\rm int} - \tfrac{\sf k}{4} w^2-a\big) \ .
\ee
Then recalling that
\be
J^3_0=\t J^3_0+ \tfrac{{\sf k}w}{2} \ ,
\ee
we find that the eigenvalue of $J^3_0$ is
\be
m = \t m+ \tfrac{{\sf k}w}{2} =  {1\over w}\big( -\tfrac{j(j-1)}{{\sf k}-2} +N^{\rm AdS} + h_0 + N^{\rm int} -a\big) +\tfrac{{\sf k}w}{4} \ .
\ee
In particular, now $N^{\rm AdS}$ and $N^{\rm int}$ are not bounded, and thus we can get a proper string spectrum in this manner.

\subsubsection{The supersymmetric case}\label{sec:susyws}

The analysis for the supersymmetric case is similar, although the details are a bit different. For generic level ${\sf k}$ one can formulate the theory in the NS-R description, i.e.\ very similarly to the bosonic description above, and the only difference is that one has to replace the bosonic current algebras $\mathfrak{sl}(2,\mathds{R})_{\sf k}$ by their ${\cal N}=1$ superconformal extensions. However, for ${\sf k}=1$, this description breaks (somewhat) down, and it is more convenient to work instead with the hybrid formalism introduced in \cite{Berkovits:1999im}, where the ${\rm AdS}_3\times {\rm S}^3 $ part of the background  is captured by a WZW model based on the affine super Lie algebra $\mathfrak{psu}(1,1|2)_{\sf k}$ \cite{Eberhardt:2018ouy}. Its bosonic subalgebra is $\mathfrak{sl}(2,\mathds{R})_{\sf k} \oplus \mathfrak{su}(2)_{\sf k}$, and we denote the corresponding generators by $J^a_n$ and $K^a_n$, respectively, where the $\mathfrak{su}(2)_{\sf k}$ currents satisfy 
\begin{align}
{}[ K^3_m,K^3_n] & = \tfrac{\sf k}{2} m \delta_{m,-n} \\
{}[K^3_m,K^\pm_n] & = \pm K^\pm_{m+n} \\
{}[K^+_m,K^-_n] & = 2 K^3_{m+n} + {\sf k}\,  m \delta_{m,-n} \ . 
\end{align}
In addition the algebra $\mathfrak{psu}(1,1|2)_{\sf k}$ contains eight fermionic generators $S^{\alpha\beta\gamma}_n$, that transform as $2\cdot ({\bf 2},{\bf 2})$ with respect to this bosonic subalgebra, see e.g.\ \cite{Eberhardt:2018ouy} for the full set of relations.

At level  ${\sf k} = 1$ the superaffine algebra $\mathfrak{psu}(1,1|2)_1$ has effectively only one highest weight representation, whose generating states transform in the (continuous) $j=\frac{1}{2}$ representation of $\mathfrak{sl}(2,\mathds{R})$, as well as the usual spin $l=\frac{1}{2}$ representation of $\mathfrak{su}(2)$ \cite{Eberhardt:2018ouy}. This defines a short representation with respect to the action of the fermionic zero modes, and the other highest weight states transform in the $(j=0,l=0)$ and $(j=1, \ell=0)$ representation of the bosonic subalgebra $\mathfrak{sl}(2,\mathds{R})_1 \oplus \mathfrak{su}(2)_1$, see \cite{Eberhardt:2018ouy} for further details.

By essentially the same arguments as above, the spectrum still contains spectrally flowed representations, where spectral flow is now defined (for ${\sf k}=1$) by 
\begin{subequations}\label{wsspecflow}
\allowdisplaybreaks
\begin{align} 
J^3_n\  = & \ \tilde J^3_n + \tfrac{ w}{2} \delta_{n,0}\label{sl2shift}  \ ,\\[4pt]
J^\pm_n\  = & \ \tilde J^\pm_{n\mp w}  \ ,\\
K^3_n\  = & \ \tilde K^3_n + \tfrac{ w}{2} \delta_{n,0} \label{su2shift}  \ ,\\[4pt]
K^\pm_n\  = & \ \tilde K^\pm_{n\pm w}  \ ,\\
S^{\alpha \beta \gamma}_n\ = & \ \tilde S^{\alpha \beta \gamma}_{n+\frac{1}{2}w(\beta-\alpha)} \ ,\\
L_n \ = & \ \tilde L_n + w( \tilde K^3_n - \tilde J^3_n)  \ . \label{confshift}
\end{align}
\end{subequations}
Here $w$ is a (positive) integer, and the spectral flow acts on both $\mathfrak{sl}(2,\mathds{R})_1$ as well as $\mathfrak{su}(2)_1$ --- for $\mathfrak{su}(2)_1$ spectral flow does not introduce any new (non-highest weight) representations, so whether to spectrally flow the $\mathfrak{su}(2)_1$ factor is a matter of convention. What is nice about the above set-up is that the ${w^2\over 4}$ term in (\ref{ggone}) cancels out between the $J^a$ and $K^a$ currents, so we have only  terms linear in $w$ in (\ref{confshift}).

The main effect of this worldsheet spectral flow is that it turns conventional highest weight representations of $\mathfrak{psu}(1,1|2)_1$ (for which the spectrum of $L_0$ is bounded from below) into representations for which this is not the case, and the reason is the same as in the bosonic analysis above. The mass-shell condition in the $w$-spectrally flowed sector now takes the form 
\be\label{massshell}
L_0 = \tilde{L}_0 + w (\tilde{K}^3_0 - \tilde{J}^3_0) = 0 \ ,
\ee
where, as before, the tilde modes $\tilde{L}_0$, $\tilde{K}^3_0$ and $\tilde{J}^3_0$, denotes the modes that act on a conventional highest weight representation. The actual spacetime conformal dimension and $\mathfrak{su}(2)$ charge are then
\be\label{st hj}
h =J^3_0= \tilde{J}^3_0 + \tfrac{w}{2}  \ , \qquad q=K^3_0 = \tilde{K}^3_0 + \tfrac{w}{2}\ ,
\ee
as follows from (\ref{sl2shift}) and (\ref{su2shift}). 

Let us now apply this formalism to determine the spectrum. The only highest weight $\mathfrak{psu}(1,1|2)_1$ representation has $\tilde{L}_0=0$ on all the highest weight states, and it is convenient to take as the reference highest weight state the one with $m=\frac{1}{2}$ in the $l=\frac{1}{2}$ representation --- the other highest weight states are obtained from this by the action of the zero modes. Then the only highest weight state (with that ansatz) that satisfies the physical state condition (for all $w$) is the one for which 
\be
 \tilde{J}^3_0=\frac{1}{2} \ , ~~~\qquad \tilde{K}^3_0 = \frac{1}{2} \ .
\ee
Its spacetime conformal dimension and $\mathfrak{su}(2)$ charge are then 
\be
h  = \frac{w+1}{2} \ , \qquad q  = \frac{w+1}{2} \ ,
\ee
as follows from (\ref{st hj}), and it therefore corresponds to the (upper) BPS state in the $w$-cycle twisted sector \cite{Eberhardt:2018ouy}, see eq.~(\ref{upper R}) below. To obtain the other states in the $w$-cycle we consider descendant states, for which the mass-shell condition becomes 
\be
L_0 = N^{\rm ws} + w ( \tfrac{1}{2} + \sum_i \delta_i - \tilde{J}^3_0 ) = 0  \ ,
\ee
where $\delta_i$ denotes the $\tilde{K}^3_0$ charge of the different oscillators, while $N^{ws}$ is the total excitation number, $N^{\rm ws} = N^{\rm AdS} + N^{\rm S} + N^{\mathbb{T}^4}$. Naively, one would have expected there to be $8+8$ bosonic+fermionic excitation modes (corresponding to the eight transverse directions), but at ${\sf k}=1$, there are actually only $4+4$ such modes --- this is a consequence of the fact that 
$\mathfrak{psu}(1,1|2)_1$ has only $2+2$ excitation modes (instead of the $6+6$ one would have naively expected), which is a small level phenomenon, similar to the fact that $\mathfrak{su}(2)_1$ only has one oscillator, rather than the $3$ one would have naively expected. It is natural to think of these excitation modes as corresponding to the (integer moded) $\mathbb{T}^4$ directions.\footnote{In the hybrid formalism the torus is topologically twisted, and thus the fermions are integer moded. The correct identification of the `transverse' fermionic modes is somewhat subtle, but in the end, they carry charge with respect to the $\mathfrak{su}(2)$ subalgebra of $\mathfrak{psu}(1,1|2)$, see  \cite{Gaberdiel:2021njm}.} Solving the mass-shell condition for the eigenvalue of $\tilde{J}^3_0$, the corresponding spacetime charges are then
\be\label{worldsheet spectrum}
h = J^3_0 =  \frac{N^{\rm ws}}{w} + \frac{w+1}{2} + \sum_i \delta_i  \ , \qquad 
q = K^3_0 = \frac{w+1}{2} + \sum_i \delta_i \ . 
\ee
The four bosons are charge neutral ($\delta_i=0$) and only the negative modes act non-trivially. Two of the fermions have $\delta_i=-\frac{1}{2}$, while the other two have $\delta_i=+\frac{1}{2}$, and only the zero modes of the fermions with $\delta_i=-\frac{1}{2}$ (as well as all the negative modes of all fermions) act non-trivially  \cite{Eberhardt:2018ouy}. In particular, the spacetime states therefore satisfy the ${\cal N}=4$ BPS bound. As we shall explain below, see Section~\ref{sec CFT analysis}, this then reproduces precisely the single-particle spectrum of the dual symmetric orbifold CFT.

\subsection{Symmetric orbifold CFT}\label{sec symmetric orbifold}

Symmetric orbifold CFTs are constructed by taking a product of $N$ copies of a seed CFT, and orbifolding by the permutation group $S_N$, resulting in the target space
\be\label{CFT orbifold}
{\mathcal M}^N/S_N \ ,
\ee
where $\mathcal M$ is the target space of the seed CFT. In this paper, we consider the case where the seed CFT is $\mathcal M=\mathbb{T}^4$ with $\mathcal N =(4,4)$ supersymmetry. The R-symmetry is ${\rm SU}(2)_{\rm L}\times {\rm SU}(2)_{\rm R}$, where `${\rm L}$' and `${\rm R}$' denotes whether the relevant ${\rm SU}(2)$ comes from the left- or right-moving ${\cal N}=4$ superconformal algebra. If we replaced the torus $\mathbb{T}^4$ by $\mathbb{R}^4$, we would have an ${\rm SO}(4)\cong {\rm SU}(2)_1\times {\rm SU}(2)_2$ symmetry. Although this symmetry is broken upon compactification to $\mathbb{T}^4$, it still serves as a useful organising principle.

The seed CFT with $\mathcal N =(4,4)$ supersymmetry can be realised using four free bosons and four free fermions. To label the fields, we use spinor indices $A,\dot A = +,-$ for ${\rm SU}(2)_1\times {\rm SU}(2)_2$ and indices $\alpha,\bar \alpha = +,-$ for ${\rm SU}(2)_{\rm L}\times {\rm SU}(2)_{\rm R}$. 
Then the four bosons can be grouped into $X_{A\dot A}$, while the left- and right-moving fermions organise themselves as $\psi^{\alpha A}$ and $\bar \psi^{\bar \alpha A}$ respectively. (We will denote right-movers by a bar in the following.)

Just like any orbifold theory, the symmetric orbifold consists of an untwisted sector, as well as twisted sectors that are associated to the conjugacy classes of the orbifold group. In the following we shall concentrate on the `single' particle states that are associated to the twisted sectors consisting of a single cycle of length $\ell$.\footnote{The conjugacy classes of $S_N$ are labelled by partitions of $N$. The single cycle sectors we are interested in correspond to the partitions of the form $N= \ell + 1 + 1 + \cdots +1$.} If the symmetric orbifold is defined on a cylinder with coordinates 
\be
\infty<\tau<\infty \ ,~~~~~~0\leq \sigma <2\pi \ ,
\ee
the $\ell$'th twisted sector is characterised by the transformation  
\be\label{X b}
X^{(1)}_{A\dot A} \to X^{(2)}_{A\dot A} \to \dots \to X^{(\ell)}_{A\dot A} \to X^{(1)}_{A\dot A} \ ,     \qquad 
\hbox{as $\sigma \to \sigma + 2\pi$} \ ,
\ee
and similarly for the fermions, see below. To characterise this sector it is convenient to define a single field $X_{A\dot A}$ defined on $\sigma \in [0,2\pi \ell]$ such that 
\be
X_{A\dot A} (\tau,\sigma) = X^{(s)}_{A \dot A}\bigl(\tau,\sigma - 2\pi (s-1)\bigr) \ , \qquad \hbox{for $\sigma\in[2\pi (s-1),2\pi s]$}\ .
\ee
This can then be extended to a periodic field defined for all values of $\sigma$ by 
\be
X_{A\dot A}(\tau,\sigma+2\pi \ell) = X_{A\dot A}(\tau,\sigma) \ .
\ee
Its modes can thus be defined as 
\be\label{alpha}
\alpha_{A\dot A, -\frac{n}{\ell}} = \frac{1}{2\pi}\int^{2\pi \ell}_{\sigma =0} dw \, e^{-\frac{n}{\ell}w}\, \partial X_{A\dot A}(w) \ ,
\ee
where $w=\tau+i\sigma$. This mode increases the conformal dimension by $h = \frac{n}{\ell}$. We can similarly define the right-moving modes $\bar \alpha_{A\dot A, -\frac{n}{\ell}}$. 

For the fermions the analysis is a bit more subtle since one has to distinguish whether the fermions are in the Ramond or Neveu-Schwarz sector. For the following it will be convenient to work in the Ramond sector for which the fermions are periodic on the cylinder, i.e.\ for which the analogue of (\ref{X b}) is 
\be\label{psi b}
\psi^{\alpha A (1)} \to \psi^{\alpha A (2)} \to \dots \to \psi^{\alpha A (\ell)} \to  \psi^{\alpha A (1)}  \ ,     \qquad 
\hbox{as $\sigma \to \sigma + 2\pi$} \ .
\ee
Similar to the bosons above, we can define a single fermion field on $\sigma \in [0,2\pi \ell]$ such that 
\be
\psi^{\alpha A} (\tau,\sigma) = \psi^{\alpha A (s) }\bigl(\tau,\sigma - 2\pi (s-1)\bigr) \ , \qquad \hbox{for $\sigma\in[2\pi (s-1),2\pi s]$}\ ,
\ee
which then satisfies 
\be
\psi^{\alpha A} (\tau,\sigma+2\pi \ell) = \psi^{\alpha A}(\tau,\sigma) \ .
\ee
As a consequence, its modes are of the form 
\be\label{d}
d^{\alpha A}_{-\frac{n}{\ell}} = \frac{1}{2\pi i}\int^{2\pi \ell}_{\sigma =0} dw \, e^{-\frac{n}{\ell}w}\psi^{\alpha A}(w) \ ,
\ee
where $w=\tau+i\sigma$. Here $n$ is an integer and the mode increases the conformal dimension by $h = \frac{n}{\ell}$. The corresponding modes for the right-moving fermions will be denoted by $\bar d^{\bar \alpha A}_{-\frac{n}{\ell}}$.

The Ramond ground state in the $\ell$-cycle twisted sector has a 16-fold degeneracy due to the presence of four fermionic zero modes both for the left-movers and for the right-movers. The state that is annihilated by 
\be
d_0^{-A}\,  |0^{--}_\ell\rangle_{\rm R} = \bar{d}_0^{-A}\,  |0^{--}_\ell\rangle_{\rm R} = 0 \ , \qquad (A = \pm)
\ee
has then the charges 
\be\label{lower R}
|0^{--}_\ell\rangle_{\rm R} \ : \qquad h^{\rm R}=\bar h^{\rm R} = \frac{\ell}{4} \ , ~~~~    q^{\rm R}=\bar q^{\rm R}=-\frac{1}{2}\ .
\ee
For the following, it will also be convenient to introduce the `top' state in the Clifford representation, 
\be\label{upper R}
|0^{++}_\ell\rangle_{\rm R} =  d^{++}_0 d^{+-}_0 \bar d^{++}_0 \bar d^{+-}_0 |0^{--}_\ell\rangle_{\rm R}\ : \qquad h^{\rm R}=\bar h^{\rm R} = \frac{\ell}{4} \ , ~~~~ q^{\rm R}=\bar q^{\rm R}=\frac{1}{2}  \ .
\ee
The state $|0^{--}_\ell\rangle_{\rm R}$ generates the full Fock space by the action of the left- and right-moving modes 
\be\label{lower R modes}
|0^{--}_\ell\rangle_{\rm R} \ : \quad  \alpha_{A\dot A, -\frac{n}{\ell}}\ , ~~  \bar{\alpha}_{A\dot A, -\frac{n}{\ell}}\ , ~~
d^{\alpha A}_{-\frac{n}{\ell}} \ , ~~ \bar{d}^{\bar \alpha A}_{-\frac{n}{\ell}} \ , \ \ (n>0)  \quad 
 \text{and} \quad  d^{+A}_{0}\ , ~~  \bar{d}^{+A}_{0} \ ,
\ee
while for $|0^{++}_\ell\rangle_{\rm R}$ the corresponding statement is 
\be\label{upper R modes}
|0^{++}_\ell\rangle_{\rm R} \ : \quad  \alpha_{A\dot A, -\frac{n}{\ell}}\ , ~~  \bar{\alpha}_{A\dot A, -\frac{n}{\ell}}\ , ~~
d^{\alpha A}_{-\frac{n}{\ell}} \ , ~~ \bar{d}^{\bar \alpha A}_{-\frac{n}{\ell}} \ , \ \ (n>0)  \quad 
 \text{and} \quad  d^{-A}_{0}\ , ~~  \bar{d}^{-A}_{0} \ .
\ee

\subsubsection{${\cal N}=4$ spectral flow}

The $\mathcal N =4$ superconformal algebra possesses an automorphism that is also usually referred to as `spectral flow'; in order to distinguish it from the worldsheet spectral flow automorphism, see eq.~(\ref{wsspecflow}), we shall refer to it as  $\mathcal N =4$ spectral flow. It acts on the ${\cal N}=4$ generators as 
\begin{subequations}\label{N=4 flow}
\begin{align}
{J}^\pm_n & = \tilde{J}^{\pm}_{n\pm \eta}  \ , \\[2pt] 
{J}^3_n & = \tilde{J}^3_n + \frac{\eta c}{12} \, \delta_{n,0}  \ , \label{N4sfj}\\ 
{L}_n & = \tilde{L}_n + \eta  \tilde{J}^3_n + \frac{\eta^2 c}{24} \, \delta_{n,0}   \ ,\label{N4sfh}\\ 
{G}^{\beta}{}_{\dot A, n} & = \tilde{G}^\beta{}_{\dot A, n+ \frac{1}{2} \eta \beta} \ ,
\end{align}
\end{subequations}
where the $G^\beta{}_{\dot A}$ are the supercurrents with $\beta = \pm$, while $\eta$ describes the `amount' of spectral flow.  If $\eta\in \mathbb{Z}$ is odd, this $\mathcal N =4$ spectral flow maps the Ramond sector to the Neveu-Schwarz sector (in which the supercurrents are half-integer moded) and vice versa. In terms of the individual free boson and fermion modes, the bosonic modes are unaffected by this $\mathcal N =4$ spectral flow (as they are uncharged with respect to the R-symmetry), while the fermions transform as 
\be\label{flow mode}
 {d}^{\pm A}_{-\frac{n}{\ell}} = \tilde{d}^{\pm A}_{- (\frac{n}{\ell}\mp \frac{\eta}{2})} \ .
\ee

\subsubsection{The spectrum of the symmetric orbifold}\label{sec:ssymspec}

With these preparations at hand, we can now describe the single particle spectrum of the symmetric orbifold theory. We shall work with the Ramond sector description of the CFT; because of the ${\cal N}=4$ spectral flow of the previous subsection, this is just a matter of convenience. 

Let us begin by explaining the ground state of the theory that arises from the untwisted sector. Given that we have fermionic zero modes in the Ramond sector, the ground state is degenerate, and we shall work with the convention that `the ground' state of the $c=6$ seed CFT is the state  $ |0^{--}_1\rangle_{\rm R}$  with the charges, see eq.~(\ref{lower R}) with $\ell=1$,
\be\label{ground}
h^0=\bar h^0={c\over 24}={1\over 4} \ , ~~~~q^0=\bar q^0 =-{1\over 2} \ .
\ee

We are interested in the single particle excitations of the symmetric orbifold, and they arise from excitations in a single $w$-cycle twisted sector (corresponding to $w$ singly wound copies of the string being combined into one $w$-fold wound copy), with all the other $(N-w)$ copies sitting in the ground state $ |0^{--}_1\rangle_{\rm R}$. It will be convenient to take the generating state of the $w$-cycle twisted sector to be $|0^{++}_{w}\rangle_{\rm R}$. The other (single particle) states in this $w$-cycle twisted sector are then obtained by acting with the modes in eq.~(\ref{upper R modes}) on this reference state, and their dimensions and charges are, see (\ref{upper R}) 
\be\label{2.50}
h^{(w)} = \frac{w}{4}  + \frac{N^{\rm CFT}}{w}\ , \qquad q^{(w)} = \frac{1}{2}  + \sum_i \delta^{\rm CFT}_i \ , 
\ee
where $N^{\rm CFT}$ is the total excitation number of the modes in units of $\frac{1}{w}$, while $\delta_j^{\rm CFT}$ is the $\mathfrak{su}(2)$ charge of each excitation.  (A similar formula also holds for the right-movers.)  Relative to the charges of $w$ copies of the ground state (\ref{ground}) the excitation energy and $\mathfrak{su}(2)$ charge is thus\footnote{Note that the remaining $(N-w)$ ground states $ |0^{--}_1\rangle_{\rm R}$ go along for the ride, and were also not included in eq.~(\ref{2.50}).}  
\be\label{dualCFT}
\Delta h = h^{(w)} - w h^0 = \frac{N^{\rm CFT}}{w} \ , \qquad \Delta q = q^{(w)} - w q^0 = \frac{w+1}{2} + \sum_i \delta^{\rm CFT}_i \ .
\ee

\subsection{Comparison of spectrum}\label{sec CFT analysis}

We are now in the position to explain that the worldsheet spectrum of Section~\ref{sec:susyws} reproduces precisely the single particle excitation spectrum we have just described. In order to understand how this works 
in detail, we first need to understand how the dual CFT charges can be calculated from the worldsheet. 

\subsubsection{The correct identification}\label{sec:iden}

Let us first analyse in some detail how the worldsheet symmetry operators are to be identified with those in the dual CFT. The left-moving contribution to the energy is the eigenvalue of the worldsheet operator $J^3_0$, and the left-moving charge of the worldsheet state is the eigenvalue of the worldsheet operator $K^3_0$. These operators correspond in the target space to the Killing vectors
\be\label{LKill}
J^3_0=\frac{1}{2}(E+P_y)= \frac{i}{2}(\partial_t-\partial_y) \ , ~~~~~
K^3_0=\frac{1}{2}(-J_\phi+J_\psi)= \frac{i}{2}(\partial_\phi-\partial_\psi) \ ,
\ee
where we are using the coordinates of eq.~(\ref{mmone}), in which the ${\rm AdS}_3$ directions are orthogonal to those associated to ${\rm S}^3$. For the right-movers the analogous statements are 
\be\label{RKill}
\bar J^3_0= \frac{1}{2}(E-P_y)= \frac{i}{2}(\partial_t+\partial_y) \ , ~~~~~
\bar K^3_0= -\frac{1}{2}(J_\phi+J_\psi)=\frac{i}{2}(\partial_\phi+\partial_\psi) \ .
\ee
It will be convenient to identify the coordinates of the dual CFT (living on the boundary of AdS$_3$) with those 
inherited from the Ramond sector coordinates of eq.~(\ref{mmtwo}). Using (\ref{mmthree}), the corresponding  Killing vectors are therefore\footnote{This transformation of charges was also studied in \cite{Balasubramanian:2000rt}, using a different approach.}
\bea
J_0^{3\,R}&\equiv& \frac{i}{2}(\partial_{t_R}-\partial_{y_R})= \frac{i}{2}\left ( \partial_t-\partial_\phi\right ) -  \frac{i}{2}(\partial_y-\partial_\psi)= J_0^3-K_0^3 \ ,\nonumber\\
K_0^{3\,R}&\equiv& \frac{i}{2}(\partial_{\phi_R}-\partial_{\psi_R})=\frac{i}{2}(\partial_\phi-\partial_\psi) = K_0^{3} \ , \label{2.54}
\eea
and similarly for the right-movers. It then follows from the values for  $J_0^3$ and $K_0^3$, as given in eq.~(\ref{worldsheet spectrum}), that 
\be
J_0^{3\,R}={N^{\rm ws}\over w} \ ,~~~~~~~
K_0^{3\,R}={w+1\over 2}+\sum_i\delta_i \ .
\ee
This now reproduces directly the dual CFT answer, see eq.~(\ref{dualCFT}), provided that $N^{\rm ws} = N^{\rm CFT}$ and $\delta_i = \delta_i^{\rm CFT}$. In fact, this identification arises if we identify each worldsheet mode with the `corresponding' mode in the dual CFT, except that the mode number is divided by $w$.

We should mention in passing that one could also work with the NS sector of the dual CFT, and identify its spectrum with that of the worldsheet theory; in effect that amounts to performing the ${\cal N}=4$ spectral flow in the dual CFT, as well as the coordinate transformation (\ref{mmthree}) in the spacetime. In fact, this is what was done in the original papers \cite{Eberhardt:2018ouy}, see also \cite{Gaberdiel:2018rqv}.

\section{Spectrum on the $({\rm AdS}_3\times {\rm S}^3)/ \mathbb{Z}_k \times \mathbb{T}^4$ orbifold}\label{sec orbifold}

So far we have described the agreement of the spectrum between string theory on ${\rm AdS}_3\times {\rm S}^3 \times \mathbb{T}^4$ and the dual symmetric orbifold CFT where the ground state comes from the untwisted sector of the symmetric orbifold. We now proceed to the main task of this paper: the analysis for the case where the bulk geometry is an orbifold of this geometry. In this section, we study the string spectrum on  $({\rm AdS}_3\times {\rm S}^3)/ \mathbb{Z}_k \times \mathbb{T}^4$, and show how its spectrum is reproduced by a certain sector of the dual symmetric orbifold. 

\subsection{The geometry of $({\rm AdS}_3\times {\rm S}^3)/ \mathbb{Z}_k \times \mathbb{T}^4$}

We have noted in section \ref{secgeometries} that the bound states of NS1 and NS5 branes produces many degenerate states. The state with the largest angular momentum gives the geometry (\ref{mmfive0}). Some other states have a form that is an orbifold of ${\rm AdS}_3\times {\rm S}^3$ in the near-horizon limit. In this section we will look at geometries which arise from extremal bound states of the NS1-NS5 system. (In the next section we will look at geometries which are also orbifolds but where the NS1-NS5 bound state is excited by momentum and anti-momentum charges $P$ and $\bar P$.)

For these states, the string metric in the near horizon limit is
\bea
ds^2_{\rm string}&=&{f\over \sqrt{Q_1Q_5}}\left[-d\t t^{\, 2}+d\t y^2\right]+\sqrt{Q_1Q_5}\Big[{d\t r^2\over \t r^2+{a^2\over k^2}}+d\t \theta^2+\cos^2\!\t \theta d\t \psi^2+\sin^2\!\t \theta d\t \phi^2\Big]\nonumber\\
&& -2{a\over k} \sin^2\!\t \theta d\t\phi d\t t-2{a\over k}\cos^2\!\t\theta d\t\psi d\t y  \ ,
\label{mmfive}
\eea
where $Q=\sqrt{Q_1Q_5}$ and $a={\sqrt{Q_1Q_5}\over R_y}$ as before and 
\be
f=\t r^2+{a^2\over k^2}\cos^2\!\t \theta \ .
\ee
Here $k$ is a positive integer, and the case $k=1$ brings us back to the metric of eq.~(\ref{mmfive0}). 
We write 
\be\label{R coordinates}
t_R={\t t\over R_y} \ , ~~~y_R={\t y\over R_y} \ , ~~~r_R={\t r\over a} \ , ~~~\theta_R=\t \theta \ , ~~~\psi_R=\t\psi \ , ~~~\phi_R=\t\phi \ ,
\ee
to find
\bea
ds^2_{\rm string}&=&Q\bigg [(r_R^2+{1\over k^2}\cos^2\!\theta_R)(-dt_R^2+dy_R^2) \nonumber \\ 
& & \qquad +\Big({dr_R^2\over r_R^2+{1\over k^2}}+d\theta_R^2+\cos^2 \!\theta_R d\psi_R^2+\sin^2\!\theta_R d\phi_R^2\Big) \\
&& \qquad -2{1\over k} \sin^2\!\theta_R d \phi_R dt_R-2{1\over k}\cos^2\!\theta_R d\psi_R dy_R\bigg ] \ . \nonumber
\label{Rmetric}
\eea
This metric has cross terms between the coordinates, but we can eliminate them by defining
\be
\psi = \psi_R  - {1\over k} {y_R} \ , ~~~\phi = \phi_R - {1\over k}{t_R}  \ .
\ee
The  metric then becomes
\be
ds^2_{\rm string}=Q\big[-( r_R^2 + {1\over k^2}) dt_R^2 + {d r_R^2\over  r_R^2 +{1\over k^2}} +  r_R^2 dy_R^2 + d \theta_R^2 +\cos^2\!\theta_R d \psi^2 +\sin^2\!\theta_R d\phi^2\big] \ .
\ee
Finally, with the definitions 
\be\label{Rcc}
 t={ t_R\over k} \ , ~~~ y= {y_R\over k} \ , ~~~ r={k r_R} \ , ~~~\theta=\theta_R \ , ~~~\psi = \psi_R  - {1\over k} {y_R} \ , ~~~\phi = \phi_R - {1\over k}{t_R}  \ ,
\ee 
this can be rewritten as 
\be
ds^2_{\rm string}=Q\big[-(  r^2 +1) d t^2 + {d r^2\over   r^2 +1} +   r^2 d y^2 + d \theta^2 +\cos^2\!\theta d \psi^2 +\sin^2\!\theta d \phi^2\big] \ .
\label{mmten}
\ee
In the Ramond sector coordinates of eq.~(\ref{R coordinates})  we have the identifications
\be
(y_R, \psi_R, \phi_R)\sim (y_R+2\pi, \psi_R, \phi_R)\sim (y_R, \psi_R+2\pi, \phi_R)\sim (y_R, \psi_R, \phi_R+2\pi) \ .
\ee
Thus (\ref{mmten}) is the metric of global ${\rm AdS}_3\times {\rm S}^3$ but with the orbifold identifications
\be
(y,\psi, \phi) \sim \bigl(y+\tfrac{2\pi}{k}, \psi - \tfrac{2\pi}{k} , \phi) \sim (y, \psi +2\pi , \phi \bigr)
 \sim (y, \psi  , \phi +2\pi) \ .
\ee

\subsection{Worldsheet analysis}

Next we want to understand how to describe this geometry from the worldsheet perspective. To this end we think of the geometry as arising via a $\mathbb{Z}_k$ orbifold acting on the coordinates
\be\label{spacetime Zk}
(y,\psi,\phi)\sim \bigl(y+\tfrac{2\pi}{k},\psi-\tfrac{2\pi}{k},\phi\bigr) \ .
\ee
In terms of the worldsheet fields and using the identification of eqs.~(\ref{LKill}) and (\ref{RKill}), 
this then corresponds to the orbifold action
\be\label{worldsheet Zk}
\mathbb{Z}_k: \quad g= e^{\frac{2\pi i}{k}\,  (J^3_0 - K^3_0)} \otimes e^{-\frac{2\pi i}{k} (\bar{J}^3_0 - \bar{K}^3_0)}  \ , 
\ee
where $J^3_0$ and $K^3_0$ are the Cartan generators of the left moving $\mathfrak{sl}(2,\mathds{R})$ and $\mathfrak{su}(2)$, respectively, while $\bar J^3_0$ and $\bar K^3_0$  are the corresponding generators for the right movers. 

As with any orbifold, the worldsheet theory will thus consist of an untwisted sector, consisting of these states of the original worldsheet theory that are invariant under the orbifold action, i.e.\ under (\ref{worldsheet Zk}). In addition, there will be `new' twisted sectors describing strings that only close up to the action by  a non-trivial element of the orbifold group $\mathbb{Z}_k$, i.e.\ that satisfy 
\begin{align}
& t(\tau,\sigma+2\pi) = t(\tau, \sigma)  \ , \n
& y(\tau,\sigma+2\pi) = y(\tau,\sigma) + 2\pi \tfrac{n}{k} \,    \n 
& \psi(\tau,\sigma+2\pi) = \psi(\tau,\sigma) - 2\pi \tfrac{n}{k} \label{spectral1} \ .
\end{align}
Comparing to eq.~(\ref{wsspecflow}), we see that the first two lines are correctly accounted for by the spectral flow by $w=\frac{n}{k}$ of Section~\ref{sec:bosspectral}. Combining with the action of the fermions, i.e.\ by adding the $K^3$ terms to spectral flow, compare eq.~(\ref{ggone}) and (\ref{wsspecflow}), then also produces correctly the third line of (\ref{spectral1}). Thus the $n$'th twisted sector of our $\mathbb{Z}_k$ orbifold theory is simply described by the spectrally flowed sectors, for which spectral flow acts via (\ref{wsspecflow}) and affects simultaneously left- and right-movers, except that now $w=\frac{n}{k}+\mathbb{Z}$, where $n=0,\ldots,k-1$ see also \cite{Martinec:2001cf,Son:2001qm}. In fact, this follows also quite abstractly from the fact that $\frac{1}{k}(J^3_0-K^3_0)$ is the exponent of the orbifold group element, see Appendix~\ref{app WZW orbifold} for a simpler incarnation of the same idea. 

The analysis of the physical spectrum thus proceeds very analogously to what was done in Section~\ref{sec:susyws}, and we still have eq.~(\ref{worldsheet spectrum}) 
\be\label{worldsheet spectrum f}
h =  J^3_0  = \frac{N^{\rm ws}}{w} + \frac{w+1}{2} + \sum_i \delta_i \ , \qquad q = K^3_0  = \frac{w+1}{2}+  \sum_i \delta_i \ ,
\ee
except that now $w$ takes the values $w=\frac{n}{k}$, where $n$ is a positive integer. Similar expressions hold for the right movers. 

In each twisted sector we furthermore have to impose the orbifold invariance condition, i.e.\ that the twisted sector state is invariant under the action of (\ref{worldsheet Zk}). Since all the oscillators are eigenvectors under $J^3_0$ and $L^3_0$, this then leads to the condition that 
\be\label{orbinv}
(h - q) - (\bar{h} - \bar{q}) \in k \cdot \mathbb{Z} \ . 
\ee
Here $(h,\bar{h})$ are the conformal dimensions of the corresponding spacetime state, while $(q,\bar{q})$ are the eigenvalues under the $\mathfrak{su}(2)$ Cartan generators of the ${\cal N}=4$ superconformal algebra. 

\subsection{Dual CFT analysis}\label{sec proposal}

Next we want to identify the corresponding states in the dual symmetric orbifold CFT, following the analysis in Section~\ref{sec CFT analysis}. While (\ref{LKill}) and (\ref{RKill}) remain true as before, the dictionary to the Ramond sector coordinates is now different, compare eq.~(\ref{mmthree}) to (\ref{Rcc}), and hence $J^{3\,R}_0$ and $K_0^{3\,R}$ now differ from (\ref{2.54}) and are instead given by 
\bea
J_0^{3\,R}&\equiv& \frac{i}{2}(\partial_{t_R}-\partial_{y_R})={i\over 2k} \left ( \partial_t-\partial_\phi\right ) - {i\over 2k} (\partial_y-\partial_\psi)={1\over k} J_0^3-{1\over k} K_0^3 \ ,\nonumber\\
K_0^{3\,R}&\equiv&\frac{i}{2}( \partial_{\phi_R}-\partial_{\psi_R})=\frac{i}{2}(\partial_\phi-\partial_\psi) = K_0^3 \ . \label{3.16}
\eea
Substituting (\ref{worldsheet spectrum f}) for $J_0^3$ and $K_0^3$ we therefore find
\be\label{3.17}
J_0^{3\,R}={N^{\rm ws}\over wk} \ , \qquad\qquad
K_0^{3\,R}={w+1\over 2}+\sum_i\delta_i \ .
\ee
We now postulate that the dual CFT states are the single particle excitations around the new ground state $(
|0^{--}_k\rangle_{\rm R})^{N/k}$, i.e.\ the state where always $k$ copies of the string are wound together.\footnote{Formally, we should therefore assume that $N$ is divisible by $k$, but since we are anyway working in the large $N$ limit, this is irrelevant.} Each such ground state component has dimension and charge
\be\label{groundk}
h^0=\bar h^0={c\over 24}={6k\over 24}={k\over 4} \ , ~~~~q^0=\bar q^{\, 0} =-{1\over 2} \ .
\ee
In order to understand what the single particle excitations should be, let us recall that 
in the theory before orbifolding, these excitations were described by twisting together $w$ of the singly wound copies describing the vacuum spacetime, to obtain a component string with twist $w$. The analogue of this procedure now is to take $w$ of the $k$-twisted copies and join them together to make a single multi-wound string with winding $wk$. (The other $N-wk$ copies of the $c=6$ CFT remain twisted in sets with winding $k$ each.) As before, we take the ground state of this $w$-cycle excitation to be the state 
$|0^{++}_{wk}\rangle_{\rm R}$, and then allow arbitrary oscillator excitations (\ref{upper R modes}) on top of it (subject to some orbifold invariance condition to be discussed below). Note that the excitation energies on the multi-wound component string with winding $wk$ come in units of ${1\over wk}$. Let $N^{\rm CFT}$ be the total dimension of all oscillator excitations in units of ${1\over wk}$, while $\delta_i^{\rm CFT}$ are the ${\rm SU}(2)_L$ charges  of these excitations. Then the excited string with winding $wk$ has left moving dimensions and charges -- again, we have not included the dimensions and charges of the remaining $\frac{N-wk}{k}$ $|0^{--}_k\rangle_{\rm R}$ ground states
\be\label{hqex}
h^{(wk)}={wk\over 4} + \frac{N^{\rm CFT}}{wk} \ , ~~~q^{(wk)}={1\over 2}  +\sum_i \delta_i^{\rm CFT} \ .
\ee
Relative to the $w$ ground states with dimension and charge (\ref{groundk}), the excitation energy and charge is therefore 
\be
\Delta h = h^{(wk)} - w h^0 = {N^{\rm CFT}\over wk} \ , ~~~\Delta q = q^{(wk)} - w q^0 =  {w+1\over 2}+\sum_i \delta_i^{\rm CFT} \ ,
\ee
thereby reproducing exactly eq.~(\ref{3.17}), provided we make the identifications 
 \be
N^{\rm ws} = N^{\rm CFT}  \ , \qquad \delta_i = \delta_i^{\rm CFT} \ . 
\ee

We have so far been a bit vague on what values $w$ takes, but also from the symmetric orbifold perspective it is now natural to consider the case where $w=\frac{n}{k}$, since only $wk$ needs to be an integer for the above analysis to make sense. In particular the $n$'th twisted sector of the worldsheet theory describes the single particle excitation where one $k$-wound string splits up into an $wk=n$-wound string (and a $(k-n)$-wound string), and we consider arbitrary excitations in the $n$-wound copy. Finally, the orbifold invariance condition that restricts the set of allowed single particle excitations in the symmetric orbifold exactly agrees with eq.~(\ref{orbinv}).

\section{More general ${\rm AdS}_3\times {\rm S}^3$ orbifolds}\label{sec:genorb}

The generalisation to a more general class of orbifold geometries is now straightforward, and we shall therefore be somewhat brief. Orbifolds of this kind were obtained for NS1-NS5-P charges and for the more general case of NS1-NS5-P-${\bar {\rm P}}$ in \cite{Jejjala:2005yu,Giusto:2012yz,Chakrabarty:2015foa}. The metric for this larger class has the form
\bea
ds^2_{\rm string}&=&-\frac{1}{Q}\big(\t r^2+{a^2\over k^2}\big)d\t t^2 
+ \frac{\t r^2}{Q} d\t y^2
+Q\bigg [{d\t r^2\over \t r^2+{a^2\over k^2}}+d\t \theta^2\nonumber \\
&&~~~~+\cos^2\!\t \theta \Big(d\t\psi-{s+\bar s+1\over k R_y} d\t y+{s-\bar s\over k R_y} d\t t\Big)^2 \label{mmsevent0} \\
&&~~~~+\sin^2\!\t \theta \Big(d\t \phi+{s-\bar s\over k R_y}d\t y-{s+\bar s+1\over k R_y} d\t t\Big)^2\,\bigg ] \ ,\nonumber 
\eea
where $s$ and $\bar s$ are integers. We define the scaled variables
\be
t_R={\t t\over R_y} \ , ~~~y_R={\t y\over R_y} \ , ~~~r_R={\t r\over a} \ , ~~~\theta_R=\t \theta \ , ~~~\psi_R=\t\psi \ , ~~~\phi_R=\t\phi \ .
\ee
Then the metric in the Ramond sector coordinates is
\bea
ds^2_{\rm string}&=&Q\bigg [-( r_R^2+{1\over k^2})d t_R^2+{d r_R^2\over  r_R^2+{1\over k^2}}+ r_R^2 d y_R^2+d \theta_R^2\nonumber \\
&&~~~~+\cos^2\! \theta_R \Big(d\psi_R-{s+\bar s+1\over k} d y_R+{s-\bar s\over k} d t_R\Big)^2 \label{mmsevent} \\
&&~~~~+\sin^2\! \theta_R \Big(d \phi_R+{s-\bar s\over k}d y_R-{s+\bar s+1\over k} d t_R\Big)^2\,\bigg ] \ . \nonumber
\eea
We furthermore define new coordinates to remove the cross terms and bring the radial coordinate to a natural form via
\bea
&&t={ t_R\over k} \ , ~~~ y= {y_R\over k} \ , ~~~r=k r_R \ , ~~~\theta=\theta_R \ ,\nonumber \\
&& \psi=\psi_R-{s+\bar s+1\over k} y_R+{s-\bar s\over k} t_R \ , ~~~\phi=\phi_R+{s-\bar s\over k}y_R-{s+\bar s+1\over k} t_R  \ ,\nonumber \\
\label{Rcc1}
\eea
so that 
\be\label{ads metric}
ds^2_{\rm string}=Q\big[-( r^2 +1) d t^2 + {d r^2\over  r^2 +1} +  r^2 d y^2 + d \theta^2 +\cos^2\!\theta d \psi^2 +\sin^2\!\theta d\phi^2\big] \ .
\ee
The identifications are then
\be\label{spacetime Zk g}
(y,\psi, \phi)\sim \bigl( y+\tfrac{2\pi}{k}, \psi - \tfrac{2\pi(s+\bar s +1)}{k}, \phi + \tfrac{2\pi (s-\bar s)}{k} \bigr)
\sim  ( y, \psi +2\pi , \phi )
\sim ( y, \psi  , \phi +2\pi) \ .
\ee

\subsection{Worldsheet analysis}
The corresponding $\mathbb{Z}_k$ orbifold action on the worldsheet is now given by 
\be\label{worldsheet g Zk}
\mathbb{Z}_k: \quad g= e^{\frac{2\pi i}{k}\,  (J^3_0 -(2s+1) K^3_0)} \otimes e^{-\frac{2\pi i}{k} (\bar{J}^3_0 -(2\bar s+1) \bar{K}^3_0)}  \ ,
\ee
since this reproduces the (first) identification in (\ref{spacetime Zk g}), using (\ref{LKill}) and (\ref{RKill}).  
Now the $n$'th twisted sector describes solutions that only close up to the action of $g^n$; in terms of the worldsheet description, the corresponding representations are then described by the modified spectral flow 
\begin{subequations}\label{specfgen}
\begin{align}
J^3_n\  = & \ \tilde J^3_n + \tfrac{ w}{2} \delta_{n,0} \ , \\[4pt]
J^\pm_n\  = & \ \tilde J^\pm_{n\mp w} \ ,\\ 
K^3_n\  = & \ \tilde K^3_n + \tfrac{ w (2s+1)}{2} \delta_{n,0} \ ,\\[4pt]
K^\pm_n\  = & \ \tilde K^\pm_{n\pm w (2s+1)} \ ,\\
S^{\alpha \beta \gamma}_n\ = & \ \tilde S^{\alpha \beta \gamma}_{n+\frac{1}{2}w((2s+1)\beta-\alpha)} \ ,\\
L_n \ = & \ \tilde L_n + w\bigl( (2s+1) \tilde K^3_n - \tilde J^3_n\bigr) + w^2 s(s+1) \, \delta_{n,0}  \  ,\label{L asy flow}
\end{align}
\end{subequations}
where now $w=\frac{n}{k}$. As before, this follows either geometrically, or using the arguments of Appendix~\ref{app WZW orbifold}. Note that, for $s=0$, this reduces to (\ref{wsspecflow}) as it should, and that for $s\neq 0, -1$, there is a central term in the spectral flow of $L_n$ since in general the central terms coming from $\mathfrak{sl}(2,\mathds{R})$ and $\mathfrak{su}(2)$ do not cancel against each other any more. Since this worldsheet orbifold action is asymmetric, we have to worry about level matching 
\cite[eq.~(2.4)]{Narain:1986qm}, see Appendix~\ref{app:AdS3S3} for more details. The level matching condition ensures that the  orbifold generator $g$ continues to have order $k$ in each twisted sector. Given the definition of $g$, see eq.~(\ref{worldsheet g Zk}), level matching reduces in our case to the requirement that the entire spectrum satisfies, see eq.~(\ref{levelmatching}) 
\be
\left(w (2s+1) \tilde K^3_0 + w^2 s(s+1)\right)- \left(w (2\bar s+1)  \tilde{\bar{ K}}^3_0 + w^2 \bar s(\bar s+1)\right)   \in \, \frac{1}{k} \, \mathbb{Z} \ ,
\ee
where $w=\frac{n}{k}$. For an abelian orbifold, the most stringent conditions arise from the first twisted sector $n=1$, resulting in
\be
\frac{s(s+1)-\bar s(\bar s+1)}{k} + (2s+1)  \tilde K^3_0 - (2\bar s+1) \tilde{{\bar K}}^3_0  \in \mathbb{Z} \ .
\ee
Since both $ \tilde K^3_0$ and $ \tilde{\bar K}^3_0$ are either both integers or both half-integers, and since both $2s+1$ and $2\bar s+1$ are odd integers, the expression $(2s+1)  \tilde K^3_0 - (2\bar s+1) \tilde{{\bar K}}^3_0$ is always an integer. 
Therefore, we obtain the level matching condition for the orbifold (\ref{worldsheet g Zk})
\be\label{level c}
 \frac{s(s+1)-\bar s(\bar s+1)}{k}  \in \mathbb{Z} \ .
\ee
\smallskip

\noindent The mass-shell condition $L_0=0$ now takes the form 
\be
 N^{\rm ws} + w\big((2s+1) (\tfrac{1}{2}+\sum_i \delta_i) - \tilde J^3_0\big) +  w^2 s(s+1) = 0 \ ,
\ee
where $N^{\rm ws}$ denotes the total excitation number before the spectral flow, while $\delta_i$ are the $\mathfrak{su}(2)$ charges of the different excitations. Solving for $\tilde J^3_0$ now gives 
\be
\tilde J^3_0= \frac{N^{\rm ws}}{w} + \frac{2s+1}{2} + (2s+1) \sum_i \delta_i   + w s(s+1) \ ,
\ee
and thus the spacetime conformal dimension equals 
\be\label{wss}
h=J^3_0 = \tilde J^3_0 + \frac{w}{2}= \frac{N^{\rm ws}}{w} + \frac{w+2s+1}{2} + (2s+1)\sum_i \delta_i + w s(s+1) \ ,
\ee
while the $\mathfrak{su}(2)$ charge is given by
\be\label{wsj}
q=K^3_0 =  \tilde K^3_0 + \frac{ w(2s+1)}{2}  = \frac{w(2s+1)+1}{2}+\sum_i \delta_i  \ .
\ee

\subsection{Dual CFT analysis}

In order to relate this to the dual CFT we proceed as in Section~\ref{sec proposal}. The generalisation of (\ref{3.16}) is 
\bea
J_0^{3\,R}&\equiv& \frac{i}{2}(\partial_{t_R}-\partial_{y_R})={i\over 2 k} \left ( \partial_t-\partial_y\right ) - {i(2 s+1)\over2 k} (\partial_\phi-\partial_\psi)={1\over k} J_0^3-{2 s+1\over k} K_0^3 \ ,\nonumber\\
K_0^{3\,R}&\equiv& \frac{i}{2}(\partial_{\phi_R}-\partial_{\psi_R})=\frac{i}{2}(\partial_\phi-\partial_\psi) = K_0^3 \ ,
\eea
as follows from (\ref{Rcc1}). 
Substituting (\ref{wss}) and (\ref{wsj}) for $J_0^3$ and $K_0^3$ we therefore find
\bea\label{hq g}
J_0^{3\,R}={N^{\rm ws}\over wk}-{ws(s+1)\over k} \ ,~~~~~~~~~~
K_0^{3\,R}={w(2s+1)+1\over 2}+\sum_i\delta_i \ .
\eea
To identify the corresponding states in the symmetric orbifold CFT we first make a proposal for what the ground state describing the metric of the near horizon geometry (\ref{mmsevent}) without any excitations in the Ramond sector coordinates should correspond to. We claim that it is described by the state  $\bigl(|\Psi_k\rangle_{s,\bar s} \bigr)^{N/k}$, where $|\Psi_k\rangle_{s,\bar s}$ is the fermionic descendant of $|0^{--}_k\rangle_{\rm R}$  \cite{Chakrabarty:2015foa} 
\begin{align}\label{bs a}
|\Psi_k\rangle_{s,\bar s} = \ & d^{-+}_{-\frac{s}{k}}d^{--}_{-\frac{s}{k}}\dots 
d^{-+}_{-\frac{2}{k}}d^{--}_{-\frac{2}{k}}d^{-+}_{-\frac{1}{k}}d^{--}_{-\frac{1}{k}} \n
&\ \times \bar d^{-+}_{-\frac{\bar s}{k}}\bar d^{--}_{-\frac{\bar s}{k}}\dots 
\bar d^{-+}_{-\frac{2}{k}}\bar d^{--}_{-\frac{2}{k}}\bar d^{-+}_{-\frac{1}{k}}\bar d^{--}_{-\frac{1}{k}}
\ |0^{--}_k\rangle_{\rm R} \ .
\end{align}
It has conformal dimension and $\mathfrak{su}(2)$ charge 
\be\label{chargesgk}
h^0=\frac{k}{4} + \frac{s(s+1)}{ k} \ , \qquad q^0=-s-\frac{1}{2} \ ,
\ee
where the ${k\over 4}$ term comes from $|0^{--}_k\rangle_{\rm R}$, while $\frac{s(s+1)}{ k} $ is the contribution of the $2s$ left-moving fermionic descendants. Similarly, $-\frac{1}{2}$ is the $\mathfrak{su}(2)$ charge of $|0^{--}_k\rangle_{\rm R}$, while $-s$ is the contribution of the $2s$ left-moving fermionic descendants. Finally, the right-moving conformal dimension and $\mathfrak{su}(2)$ charge has the same form, except that $s\mapsto \bar{s}$. It remains to check that the state is in fact part of the symmetric orbifold spectrum, i.e.\ whether it is orbifold invariant. Since the total ground state is a tensor product of $k$-cycle states, the state is orbifold invariant if and only if each $|\Psi_k\rangle_{s,\bar s}$ is. In the $k$-cycle twisted sector, on the other hand, the $S_N$ orbifold invariance simply reduces to the cyclic invariance $\mathbb{Z}_k$, and the condition that a state is orbifold invariant is equivalent to the condition that $h-\bar{h} \in \mathbb{Z}$, i.e.\ that the difference of left- and right-moving conformal dimension is an integer. Since the left-moving conformal dimension of $|\Psi_k\rangle_{s,\bar s}$ is given by the first equation of eq.~(\ref{chargesgk}), and the right-moving conformal dimension is of the same form with $s\mapsto \bar{s}$,  the orbifold invariance condition is therefore simply
\be\label{wwone}
h^0 - \bar{h}^{\, 0} = \frac{s(s+1)}{ k}  - \frac{\bar{s}(\bar{s}+1)}{ k}  \in \mathbb{Z} \ , 
\ee
i.e.\ exactly the same as (\ref{level c}). 
 
As before, the single particle excitations of this state involve a single $wk$-twisted sector, and we describe them in terms of oscillator excitations (\ref{upper R modes}) on top of the state $|0^{++}_{wk}\rangle_{\rm R}$.
By the same arguments as in (\ref{hqex}), their conformal dimension and $\mathfrak{su}(2)$ charge is now
\be\label{hqex1}
h^{(wk)}={wk\over 4} + \frac{N^{\rm CFT}}{wk} \ , ~~~q^{(wk)}={1\over 2} +\sum_i \delta_i^{\rm CFT} \ .
\ee
Relative to $w$ times the ground state energy of eq.~(\ref{chargesgk}) we therefore find for the excitation spectrum 
\begin{align}
\Delta h &=h^{(wk)} - w \, h_0=\frac{N^{\rm CFT}}{wk} -{ws(s+1)\over k} \ , \nonumber \\[2pt] 
\Delta q &= q^{(wk)} - w \, q_0={w(2s+1)+1\over 2} +\sum_i \delta_i^{\rm CFT} \ ,
\end{align}
which reproduces precisely (\ref{hq g}) provided we make the identifications $N^{\rm ws} = N^{\rm CFT}$ and $\delta_i = \delta_i^{\rm CFT} $. This argument applies irrespective of whether $w$ is an integer (i.e.\ comes from the untwisted sector), or a fraction $w=\frac{n}{k}$ (i.e.\ comes from the $n$'th twisted sector).

\section{Discussion}\label{sec:discussion}

The AdS/CFT correspondence establishes a remarkable relation between string theory on an AdS background and a dual conformal field theory. While there is obviously very strong evidence in favour of its correctness, it is difficult to get a clear understanding of the duality since it is of `strong-weak' type, i.e.\ a weakly coupled field theory is dual to strongly coupled gravity, and vice versa. 

In this paper we have explored an exact incarnation of the duality where both sides of the correspondence are under quantitative control. In particular, we have considered a generalisation of the duality relating string theory on  ${\rm AdS}_3\times {\rm S}^3\times \mathbb{T}^4$, with one unit ${\sf k}=1$ of NS-NS flux --- this is the background that is described by the bound state of $n_5=1$ NS5 branes and $n_1$ NS1 branes in the large $N=n_1$ limit --- to the symmetric orbifold of $\mathbb{T}^4$. The string theory has an exactly solvable worldsheet description using the hybrid formalism of \cite{Berkovits:1999im}, and its spacetime spectrum matches exactly that of the dual symmetric orbifold theory (which is also exactly solvable) \cite{Eberhardt:2018ouy}. More specifically, the perturbative excitations of the worldsheet theory are in one-to-one correspondence with the single-particle excitations of the dual symmetric orbifold; these consist of the states where we apply an arbitrary excitation to one $w$-cycle twisted sector, while the remaining $(N-w)$ copies are all in their single cycle ground state. Furthermore, the correlation functions of the hybrid string reproduce exactly \cite{Eberhardt:2019ywk,Dei:2020zui,Eberhardt:2020akk,Knighton:2020kuh} those of the symmetric orbifold \cite{Lunin:2000yv,Lunin:2001pw,Pakman:2009zz,Pakman:2009ab}.

As noted in the introduction, a significant amount of work has been done on string theory on ${\rm AdS}_3\times {\rm S}^3\times \mathbb{T}^4$ in the case where the number of NS5 branes is ${\sf k}>1$ \cite{Martinec:2001cf,Martinec:2002xq,Martinec:2017ztd,Martinec:2019wzw,Martinec:2018nco,Martinec:2020gkv, Bufalini:2021ndn,Martinec:2022okx,Martinec:2023zha}. In this case the worldsheet dynamics of the string can be studied through the usual NS-R formalism. Our analysis on the other hand is for the case ${\sf k}=1$, which can be accessed through the hybrid formalism that we have used. Not surprisingly, the general structure of the string orbifold and its CFT dual that we obtain is very similar to the corresponding structure obtained for ${\sf k}>1$. In particular, the orbifold consistency condition (\ref{level c}) and its CFT dual eq.~(\ref{wwone}) also arise in the case ${\sf k}>1$. Thus the main value of our analysis lies in the fact that the duality in this case is exact, and thus applies to all string states. For ${\sf k}>1$, it can be seen that the energies of stringy states do not in general agree with the corresponding dual CFT states; this is natural because the two theories are not at the same point in moduli space.

In the present paper we have considered a certain family of $\mathbb{Z}_k$ orbifolds of this set-up. String theory on  $\bigl({\rm AdS}_3\times {\rm S}^3\bigr) / \mathbb{Z}_k \times \mathbb{T}^4$ can be described, from a worldsheet perspective, by a $\mathbb{Z}_k$ orbifold of the above worldsheet theory. On the other hand, following \cite{Martinec:2001cf}, the dual CFT degrees of freedom should be accounted for as follows: we still consider the original symmetric orbifold of $\mathbb{T}^4$ but now regard the state where we have $(N/k)$ $k$-twisted sectors, each in a suitable `ground' state, as the CFT state dual to the $\mathbb{Z}_k$ orbifold of ${\rm AdS}_3\times {\rm S}^3$. (Geometrically, this means that the $N$ fundamental strings all combine into groups of $k$-wound strings, and that each such $k$-wound string is in the same state.) The `single particle' excitations of this multi-wound string configuration consist then of those states where an arbitrary excitation acts in one $wk$-cycle twisted sector, whereas the remaining copies are described by the product of the original $k$-cycle `ground' states. Our main result is that this single-particle excitation spectrum in the symmetric orbifold theory is precisely reproduced by the above worldsheet description. In particular, the case where $w\in\mathbb{Z}$ comes from the untwisted sector of the worldsheet orbifold, while the twisted sectors correspond to $k$-fractional values for $w$. 

We should stress that the energy of the state, as calculated on the worldsheet, matches the {\em excitation} energy in the symmetric orbifold, i.e.\ the difference to the energy of the reference state corresponding to the original background. Thus, as in the original description of \cite{Eberhardt:2018ouy}, the remaining unexcited $k$-cycles are invisible from the world-sheet perspective --- they have become part of the background. In particular, we can therefore  think of the orbifold geometry as arising from the condensation of these $k$-cycle twisted sector states. (With respect to the original AdS background, each single $k$-cycle describes a perturbative single particle excitation, and the background is now made up by considering $(N/k)\rightarrow \infty$ such excitations.) This general picture is also nicely in line with the ideas of \cite{Eberhardt:2020bgq,Eberhardt:2021jvj}, namely that all these backgrounds are dual to the same symmetric orbifold theory.

\begin{figure}[h]
\centering
        \includegraphics[width=8cm]{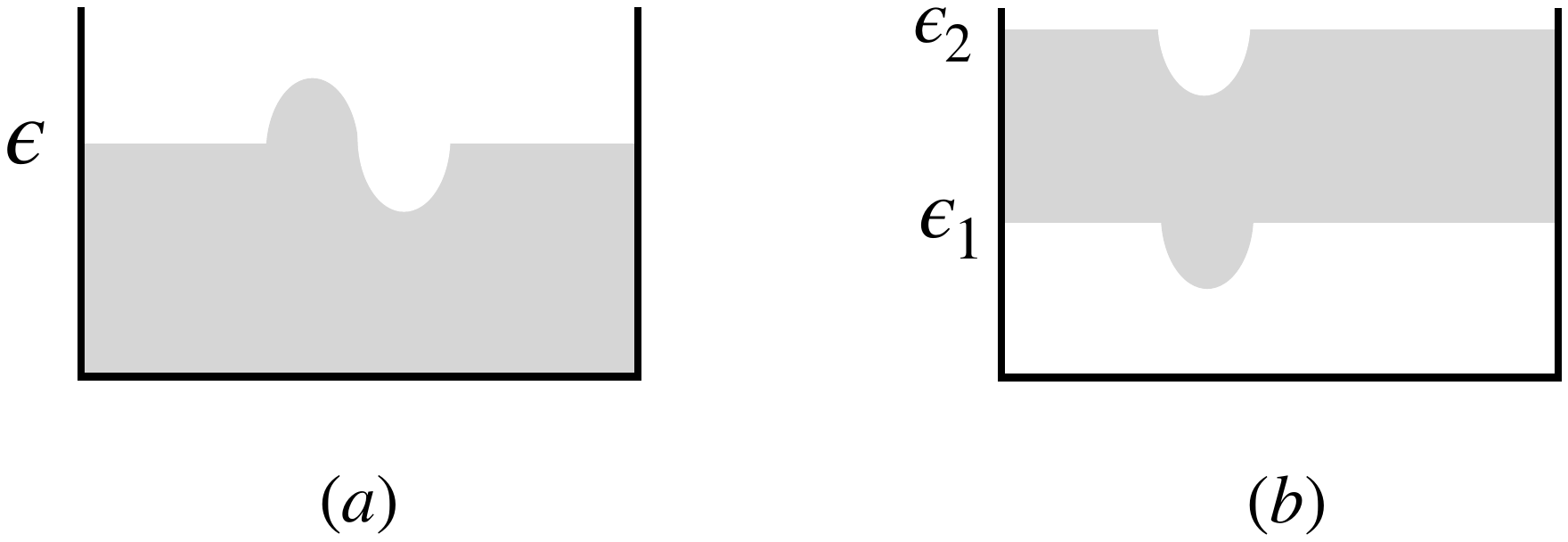}
\caption{An analogue of how AdS/CFT duality seems to be working in our situation. (a) If we describe the state of all the fermions, then this would be analogous to the BCFT description where we see all $n_1$ NS1 branes as copies of the $c=6$ seed CFT. If we describe only {\it deformations} of the Fermi sea, then this would be analogous to the gravity description where we see a string in the bulk representing the excitation. (b)  For suitable sets of occupied fermion levels, the excitation can have negative energy; we have found this situation for the energy of strings in the bulk for the case where the dual CFT  has $P-\bar P$ excitations in the unperturbed state.}
\label{fig}
\end{figure}

The picture of the AdS/CFT duality that emerges from this analysis can thus be understood through a rough analogy with the idea of a Fermi surface in a metal. In Fig.~\ref{fig}(a) we depict a Fermi sea filled up to a Fermi level $\epsilon$; an excitation on this sea is also depicted, and corresponds to changing the energy of one  of the electrons.  In one description of the system we may give the energy levels of all electrons; this is analogous to the dual CFT description where we see all $N$ copies of the seed $c=6$ CFT. But in another description we may ignore the filled sea, and just look at the changes to the Fermi surface that describe the excitation; this is analogous to the string worldsheet appearing as an excitation over the gravity background. In Fig.~\ref{fig}(b) we depict a situation where the filled Fermi levels in the metal range from $\epsilon_1$ to $\epsilon_2$. This is analogous to a {\it different} gravity background in the AdS/CFT correspondence. Note that from this perspective it is possible for an excitation to have  {\it negative} energy relative to some Fermi sea, since the electrons can occupy lower energy states than the states they occupied in the unperturbed configuration. 

\bigskip

Since some of the $\mathbb{Z}_k$ orbifolds act asymmetrically on the worldsheet CFT, there are consistency conditions, in particular, level matching, that needs to be satisfied for the worldsheet construction to make sense, see eq.~(\ref{level c}). This consistency condition also has a simple analogue in the dual CFT, where it arises as the requirement that the relevant reference state, corresponding to the unexcited background, is in fact orbifold invariant.\footnote{In \cite{Bufalini:2021ndn} a description of this condition for the case ${\sf k}\gg 1$ was given in terms of the absence of closed timelike curves and horizons.}
\smallskip

In this paper we have only compared the spectrum between the two descriptions, but it would also be interesting to show that the correlation functions match up, see also \cite{Bufalini:2022wyp,Bufalini:2022wzu} for the study of correlation functions for ${\sf k}>1$. At least structurally, this seems to be correct: if we work to leading order in $1/N$, then the worldsheet correlators should be evaluated on the sphere. Then, all intermediate states of any $n$ point function of untwisted sector states must also come from the untwisted sector. On the other hand, from the perspective of the dual CFT, the untwisted sector states of the worldsheet theory correspond to those sectors where we have $wk$-cycles (with $w\in\mathbb{N}$), while all the remaining copies are in their product $k$-cycle reference state. One can show that, to leading order in $1/N$, only sectors with $\hat{w}k$-cycles (and integer $\hat{w}\in\mathbb{N}$) can appear as intermediate states in any symmetric orbifold correlator involving external $w_i k$ cycle external states (with $w_i\in\mathbb{N}$), and this accounts nicely for the structure of the worldsheet orbifold. It will be interesting to explore this aspect further, and we hope to come back to it in due course \cite{GGM}. 

More generally, the main motivation for using the hybrid formalism of the worldsheet theory is that it allows us to obtain an exact AdS/CFT duality map. Studying such a map helps in demystifying gravity: at least at some point in moduli space, we understand all the states of the gravity theory (and their dual CFT analogues). This is important for the study of black holes and the fuzzball proposal \cite{fuzzballs,Lunin:2002iz,skenderis,mathur:2005,bw,Chowdhury:2010ct,Bena:2015bea,Bena:2016ypk,Shigemori:2022gxf,Mathur:2020ely}. We cannot hope to construct all microstates for all black holes since that would be very complicated, but we can hope to understand features of generic states by looking at simple black hole microstates. This is what the fuzzball program does, and large classes of gravity solutions have been found. These solutions can be heuristically related to the dual CFT at the orbifold point; the relation is only heuristic because the gravity solutions are at a point of moduli space where the dual CFT is strongly coupled, while the orbifold CFT is `free'. The exact AdS/CFT duality studied in the present paper takes us a little further: at least at this point in moduli space, we understand all extremal and near extremal states in the gravity theory, and we find no conceptual difference between the simple states (those whose duals have been found as fuzzballs) and more generic states (many of which have so far not been constructed). This supports the general idea of the fuzzball conjecture which states that all microstates of black holes will have the same feature observed in the simple states which have been constructed: the absence of a horizon. 

\acknowledgments

We thank Davide Bufalini, Nicolas Kovensky, Emil Martinec, and David Turton for useful discussions. MRG and SDM thank the Simons Center for Geometry and Physics for hospitality where this collaboration was begun. 
BG thanks the University of Bonn for hospitality where part of this work was done.
MRG is supported in part by the Simons Foundation grant 994306  (Simons Collaboration on Confinement and QCD Strings). The work of his group at ETH is supported by a personal grant from the Swiss National Science Foundation, as well the NCCR SwissMAP that is also funded by the Swiss National Science Foundation. BG is supported in part by the ERC Grant 787320 - QBH Structure. SDM is supported in part by the DOE grant DE-SC0011726.

\appendix

\section{The orbifold point}\label{app:orbifold}

Consider the bound state of $n_5$ NS5 branes and $n_1$  NS1 branes. The `naive' solution (i.e.\ ignoring fuzzball effects) is (in the string frame)
\bea
ds^2_{\rm string} &=& H_1^{-1}(-dt^2+dy^2)+H_5\sum_{i=1}^{4} dx_idx_i 
+ \sum_{a=1}^{4}dz_a dz_a \ , \\
e^{2\phi} &=& \frac{H_5}{H_1} \ , \\
H_1&=&1+\frac{Q_1}{r^2}~~~~~~H_5=1+\frac{Q_5}{r^2} \ . 
\eea
Here 
\be
Q_1 = \frac{g^2 \, l_s^6}{V}n_1 \ , \qquad Q_5 = l_s^2\, n_5 \ , 
\ee
where $\alpha'=l_s^2$ and the volume of the compact $T^4$ with coordinates $z_a$ is $(2\pi)^4 V$. 

First we note some general estimates. In the ${\rm AdS}$ region, the string coupling $g_{\rm AdS}$ is related to the string coupling $g$ at infinity as
\be
g_{\rm AdS}=g \, e^\phi=g\, \sqrt{Q_5\over Q_1} \ . 
\ee
The 10d Newton's constant is proportional to 
\be
G_{10}\sim g_{\rm AdS}^2 l_s^8  \sim \frac{Q_5}{Q_1}  g^2 l_s^8 
\sim {n_5\over n_1} V l_s^4 \ , 
\ee
while the size of the ${\rm S}^3$ is
\be
R_{{\rm S}^3}\sim R_{{\rm AdS}_3}   \sim \sqrt{n_5}\,  l_s \ . 
\ee
The 3d Newton's constant is then
\be
G_3\sim \frac{G_{10}}{VR_{{\rm S}^3}^3} \sim \frac{1}{\sqrt{n_5}n_1} l_s \ .
\ee
Note that $G_3\sim l_p^{(3)}$, where $l_p^{(3)}$ is the Planck length in the 3d ${\rm AdS}_3$ space. The interaction strength in this 3d gravity theory depends on the wavelength of the interacting quanta. For wavelengths $\lambda \sim R_{{\rm AdS}_3}\sim \sqrt{n_5}\, l_s$ the interaction strength is given by the dimensionless number 
\be
{G_3\over \lambda}\sim {1\over n_1n_5} \ . 
\label{kkone}
\ee
The expansion parameter of the dual CFT is 
\be
{1\over n_1n_5}\equiv {1\over N} \ .
\ee
In order to get to the orbifold point, we set $n_5=1$. Recall that if we have ${\sf k}$ NS5 branes, then each NS1 brane bound to these NS5 branes splits into ${1\over {\sf k}}$ `fractional' NS1 branes \cite{Maldacena:1996ds}. These fractional NS1 branes presumably have a complicated dynamics, but for the case ${\sf k}=1$, there is no fractionalization, and we may expect a simpler dynamics of the NS1 branes. 

Note that the gravitational interactions in this AdS region do not go to zero, since
\be
{G_3\over \lambda}\sim {1\over n_1} \ . 
\label{kkonep}
\ee
We can now take a further limit $N = n_1\rightarrow\infty$. In this limit the strings in the bulk have an interaction strength (\ref{kkonep}) which goes to zero. Thus we get free strings in the bulk. This is the limit in which we work in the present paper.

\section{The WZW model and its orbifold}\label{app WZW orbifold}

In this appendix we describe the orbifold of the $\mathfrak{su}(2)_{\sf k}$ WZW model by $\mathbb{Z}_N$ in some detail;\footnote{We denote the orbifold order by $N$ here to avoid confusion of the orbifold order (which we previously denoted by $k$) with the level of the affine algebra, ${\sf k}$.} while this is only a toy model, the key features generalise also to the theories that are discussed in this paper, see Section~\ref{app:AdS3S3}. 

At level ${\sf k}\in\mathbb{N}$, the $\mathfrak{su}(2)$ theory is rational, and there are only finitely many integrable highest weight representations of $\mathfrak{su}(2)_{\sf k}$. They are characterised by the spin $j$, and the allowed values are $j=0,\frac{1}{2},1,\ldots, \frac{{\sf k}}{2}$. The charge conjugation modular invariant (that describes the theory on the group manifold of ${\rm SU}(2)$) has then the partition function 
\be\label{WZW}
Z(\tau,\bar{\tau}) = \sum_{j=0}^{\frac{{\sf k}}{2}} \bigl| \chi_j(\tau) \bigr|^2 \ ,
\ee
where $\chi_j(\tau)$ is the specialised character associated to the spin $j$ representation ${\cal H}_j$, 
\be
 \chi_j(\tau) = {\rm Tr}_{{\cal H}_j} \bigl( q^{L_0 - \frac{c}{24}}  \bigr) \ , \qquad q = e^{2\pi i \tau} \ . 
 \ee
The specialised characters transform into one another under the modular group; in particular, we have the identity 
\be
\chi_j(-\tfrac{1}{\tau}) = \sum_{j'=0}^{\frac{{\sf k}}{2}} S_{jj'} \, \chi_{j'}(\tau) \ , 
\ee
where $S_{jj'}$ is the modular $S$-matrix of $\mathfrak{su}(2)_k$, which is explicitly given by, see e.g.\ \cite[Chapter 14.5]{DiFrancesco:1997nk} 
\be\label{Smatrix}
S_{jj'} = \sqrt{\frac{2}{{\sf k}+2}} \, \sin \Bigl( \pi \frac{(2j+1) (2j'+1)}{{\sf k}+2} \Bigr) \ . 
\ee
It is not too difficult to check that $Z(\tau,\bar{\tau}) $ in (\ref{WZW}) is modular invariant. Indeed, we have 
\begin{align}
Z(-\tfrac{1}{\tau},-\tfrac{1}{\bar{\tau}}) & =  \sum_{j=0}^{\frac{{\sf k}}{2}} \sum_{j',\bar{\jmath}'=0}^{\frac{{\sf k}}{2}} S_{jj'} \, S_{j\bar{\jmath}'} \, \chi_{j'}(\tau) \chi_{\bar{\jmath}'}(\bar\tau)  = \sum_{j'=0}^{\frac{{\sf k}}{2}}  \bigl| \chi_{j'}(\tau) \bigr|^2 \ , 
\end{align}
where we have used that the $S$-matrix is unitary, i.e.\ satisfies
\be\label{Suni}
\sum_{j=0}^{\frac{{\sf k}}{2}} S_{jj'} \, S_{j\bar{\jmath}'} = \delta_{j'\bar{\jmath}'} \ . 
\ee

\subsection{The orbifold action}

We are interested in the orbifold of this theory by $\mathbb{Z}_N$, where the fundamental generator of $g\in\mathbb{Z}_N$ acts as 
\be\label{action}
g \mapsto \exp\Bigl( \frac{2\pi i}{N} (J^3_0 + \bar{J}^{\, 3}_0 ) \Bigr) \ , 
\ee
and $J^3_0$ and $\bar{J}^{\, 3}_0$ are the left- and right-moving Cartan generators of the horizontal (zero mode) algebra. Note that the eigenvalues of $J^3_0$ and $\bar{J}^{\, 3}_0$ are separately half-integers, but that on the spectrum of (\ref{WZW}), the eigenvalues of $J^3_0 + \bar{J}^{\, 3}_0$ are always integer. Thus $g$ has order $N$ when acting on the states in (\ref{WZW}). 

\subsection{The untwisted sector}

The untwisted sector of the orbifold is obtained upon projecting the states of (\ref{WZW}) onto those that are invariant under the orbifold action. This can be implemented by inserting 
\be
\frac{1}{N} \sum_{\ell=0}^{N-1} g^{\ell}
\ee
into the trace. This leads to 
\be\label{untwisted}
Z^{(U)}(\tau,\bar{\tau}) = \frac{1}{N} \sum_{\ell=0}^{N-1} \sum_{j=0}^{\frac{{\sf k}}{2}} \bigl| \chi_j(\tfrac{\ell}{N},\tau) \bigr|^2 \ ,
\ee
where $\chi_j(\tfrac{\ell}{N},\tau)$ denotes now the unspecialised character
\be
\chi_j(z,\tau) = {\rm Tr}_{{\cal H}_j} \bigl( q^{L_0 - \frac{c}{24}} y^{J^3_0} \bigr) \ , 
\ee
with $q=e^{2\pi i\tau}$ and $y = e^{2\pi i z}$. The unspecialised characters transform as, see e.g.\ \cite[Chapter 14.5]{DiFrancesco:1997nk} or \cite[eq.~(2.9)]{Gaberdiel:2012yb}\footnote{In our conventions, $[J^3_m,J^3_n] = \frac{{\sf k}}{2} m \delta_{m,-n}$.}
 
\be
\chi_j(\tfrac{z}{\tau},-\tfrac{1}{\tau}) = \sum_{j'=0}^{\frac{{\sf k}}{2}} S_{jj'} \, e^{\pi i {\sf k} \frac{z^2}{2\tau}} \, \chi_{j'}(z,\tau) \ , 
\ee
where $S_{jj'}$ is the same matrix as above, see eq.~(\ref{Smatrix}). 

\subsection{The twisted sectors}

In order to determine the $\ell$'th twisted sector we now perform the $S$-modular transformation of the $\ell$'th term in (\ref{untwisted}); this leads to 
\begin{align}
Z^{(\ell)} (\tau,\bar{\tau}) & = \sum_{j=0}^{\frac{{\sf k}}{2}} \bigl| \chi_j(\tfrac{\ell\tau}{N\tau},-\tfrac{1}{\tau}) \bigr|^2 \\ 
& = \sum_{j=0}^{\frac{{\sf k}}{2}} \, \sum_{j',\bar{\jmath}'=0}^{\frac{{\sf k}}{2}} S_{jj'} \, S_{j\bar{\jmath}'} \, 
e^{\pi i {\sf k} \frac{\ell^2 \tau^2 }{2 N^2\tau}} \, \chi_{j'}(\tfrac{\ell \tau}{N},\tau) \, 
e^{-\pi i {\sf k} \frac{\ell^2 \bar{\tau}^2}{2 N^2\bar{\tau}}} \, \chi_{\bar{\jmath}'}(\tfrac{\ell\bar{\tau} }{N},\bar{\tau})  \\
& = \sum_{j'=0}^{\frac{k}{2}} \, \bigl| e^{\pi i {\sf k} \frac{\ell^2\tau}{2 N^2}} \chi_{j'}(\tfrac{\ell \tau }{N},\tau) \bigr|^2 \ , 
\end{align}
where we have used again (\ref{Suni}). The character that appears in the final line is simply the twisted ${\cal H}_j$ character, i.e.\ it is obtained from the original ${\cal H}_j$ character using the `spectral flow' by $\alpha=\frac{\ell}{N}$ units,\footnote{The role of the twisted affine algebra for orbifolds of WZW models was stressed in \cite{deBoer:2001nw}; other aspects of these orbifolds were also discussed in \cite{Birke:1999ik}.}  that shifts 
\begin{align}
J^3_n\  \mapsto & \ J^3_n + \tfrac{{\sf k} }{2}\, \alpha \, \delta_{n,0} \\[4pt]
J^\pm_n\  \mapsto & \ J^\pm_{n\pm \alpha} \\ 
L_n \ \mapsto & \ L_n + \alpha\, J^3_n +  \tfrac{{\sf k}}{4} \alpha^2\, \delta_{n,0} \ . \label{1.17}
\end{align}
Indeed, the exponential factor $e^{\pi i {\sf k} \frac{\ell^2\tau}{2 N^2}} = e^{2 \pi i \frac{{\sf k}}{4} \frac{\ell^2}{N^2} \tau}$ comes from the $\tfrac{{\sf k}}{4} \alpha^2\, \delta_{n,0} $ term in eq.~(\ref{1.17}), while the fact that the unspecialised character is evaluated at $z=\frac{\ell \tau}{N}$ reflects the linear term $ \alpha\, J^3_0$ in eq.~(\ref{1.17}), where $\alpha=\frac{\ell}{N}$. 
As a consequence the twisted sector has a good vector space interpretation, and it carries this spectrally flowed representation of $\mathfrak{su}(2)_{\sf k}$. It is then straightforward to impose the orbifold projection on it, simply by applying repeatedly the $T$-transformation; thus the projected $\ell$-twisted sector has the partition function 
\be
Z^{(\ell)}_{\rm inv} (\tau,\bar{\tau})  = \frac{1}{N}\, \sum_{m=0}^{N-1} 
\sum_{j'=0}^{\frac{{\sf k}}{2}} \, \bigl| e^{\pi i {\sf k} \frac{\ell^2(\tau+m)}{2 N^2}} \chi_{j'}(\tfrac{\ell (\tau+m) }{N},\tau+m) \bigr|^2 \ . 
\ee
Note that $T^N$ acts trivially since the specialised character $\chi_{j'}(z,\tau)$ is periodic in $z$ with period one, see the discussion below eq.~(\ref{action}); thus there is no `level matching anomaly' in this case. In particular, therefore, the full partition function 
\be
Z_{\rm orb} (\tau,\bar{\tau}) = Z^{(U)}(\tau,\bar{\tau})  + \sum_{\ell=1}^{N} Z^{(\ell)}_{\rm inv} (\tau,\bar{\tau}) 
\ee
is modular invariant and defines the state space of the orbifold theory.

\subsection{Orbifolds of $\mathfrak{sl}(2,\mathds{R})_{\sf k} \oplus \mathfrak{su}(2)_{\sf k}$}\label{app:AdS3S3}

Let us now sketch how the above analysis is modified if we consider an orbifold of $\mathfrak{sl}(2,\mathds{R})_{\sf k} \oplus \mathfrak{su}(2)_{\sf k}$. In order to capture the general case relevant for Section~\ref{sec:genorb}, we consider the in general asymmetric $\mathbb{Z}_N$ orbifold acting via 
\be\label{asy orbifold}
g \mapsto \exp \Bigl[ \frac{2\pi i}{N} \, \bigl( n J^3_0 + m K^3_0 + \bar{n} \bar{J}^3_0 + \bar{m} \bar{K}^3_0 \bigr) \Bigr] \ , 
\ee
where the right-movers have been denoted by a bar, and $m$, $\bar{m}$, $n$ and $\bar{n}$ are all integers.  (The case of eq.~(\ref{worldsheet g Zk}) is of this form with $n = - \bar{n}=1$, $m=-(2s+1)$, $\bar{m}=2\bar{s}+1$, and $N=k$.) The basic consistency condition of asymmetric orbifolds is the level matching condition, which requires that the orbifold projection has order $N$ in the twisted sectors, see \cite[eq.~(2.4)]{Narain:1986qm}. For example, for the first twisted sector this is simply the condition that $T^N={\bf 1}$ (where $T$ is the modular $T$-matrix), and this is equivalent to 
\be\label{levelmatching}
\frac{{\sf k}}{4N^2} (m^2 - \bar{m}^2 + \bar{n}^2 - n^2 \bigr) \in \frac{1}{N}\, \mathbb{Z} \ . 
\ee
Here we have used that the ground state energy in the $\ell$'th twisted sector is 
\begin{align}
\mathfrak{su}(2)_{\sf k}: \qquad & \delta h = \frac{{\sf k}}{4N^2} ( \ell^2 m^2) \\ 
\mathfrak{sl}(2,\mathds{R})_{\sf k}: \qquad & \delta h = - \frac{{\sf k}}{4N^2} ( \ell^2 n^2) \ , 
\end{align}
and similarly for the right-movers. The formula for $\mathfrak{su}(2)$ can be read off directly  from eq.~(\ref{1.17}); the result for $\mathfrak{sl}(2,\mathds{R})$  follows from the fact that  $\mathfrak{sl}(2,\mathds{R})_{\sf k} \cong \mathfrak{su}(2)_{- {\sf k}}$, which is the reason for the sign difference. The level-matching condition is then simply (\ref{levelmatching}), and it is not difficult to see that this is also sufficient for the other twisted sectors. For the case of eq.~(\ref{worldsheet g Zk}) this then simplifies to eq.~(\ref{level c}).

\end{document}